\documentclass[prd,preprint,amsmath,amssymb,nofootinbib,eqsecnum]{revtex4-1}
\usepackage{latexsym,graphics,color,epsfig,hyperref,mathrsfs}

\newcommand{\ie}{{i.e.}}
\newcommand{\cf}{{cf.}}

\newcommand{\viz}{{viz.}}
\newcommand{\wrt}{with respect to}
\newcommand{\lhs}{left-hand side}

\newcommand{\rhs}{right-hand side}
\newcommand{\rhss}{right-hand sides}

\newcommand{\naive}{na\"{\i}ve}

\newcommand{\be}{\begin{equation}}
\newcommand{\ee}{\end{equation}}
\newcommand{\bea}{\begin{eqnarray}}
\newcommand{\eea}{\end{eqnarray}}
\newcommand{\beas}{\begin{eqnarray*}}
\newcommand{\eeas}{\end{eqnarray*}}

\newcommand{\bear}{\begin{array}{l}}
\newcommand{\eear}{\end{array}}

\newcommand{\bcf}{\begin{center}\begin{figure}}
\newcommand{\ecf}{\end{figure}\end{center}}

\newcommand{\bct}{\begin{center}\begin{table}}
\newcommand{\ect}{\end{table}\end{center}}

\newcommand{\eq}[1]{(\ref{eq:#1})}

\newcommand{\Eqn}[1]{Equation~(\ref{eq:#1})}

\newcommand{\eqs}[2]{(\ref{eq:#1}) and~(\ref{eq:#2})}
\newcommand{\eqss}[3]{(\ref{eq:#1}), (\ref{eq:#2}) and~(\ref{eq:#3})}

\newcommand{\sect}[1]{section~\ref{sec:#1}}

\newcommand{\app}[1]{appendix~\ref{app:#1}}

\newcommand{\D}{d}
\newcommand{\Int}[1]{\int \!\! d^{\D} \! #1 \,}
\newcommand{\volume}[1]{d^{\D} \! #1 \,}
\newcommand{\MomInt}[1]{\int \!\! \frac{d^{\D} \! #1}{(2\pi)^{\D}} \,}

\newcommand{\DD}[1]{\delta^{(#1)}}

\newcommand{\der}[2]{\frac{d #1}{d #2}}

\newcommand{\pder}[2]{\frac{\partial #1}{\partial #2}}

\newcommand{\fder}[2]{\frac{\delta #1}{\delta #2}}

\newcommand{\lpartial}{\overleftarrow\partial}

\newcommand{\partialprod}[3]{D_{\underline{#1}_{#2}^{#3}}}

\newcommand{\Or}{\mathrm{O}}
\newcommand{\order}[1]{\Or \bigl( #1 \bigr)}

\newcommand{\hf}{\frac{1}{2}}
\newcommand{\smallhf}{{\textstyle \frac{1}{2}}}

\newcommand{\one}{1\!\mathrm{l}}

\newcommand{\SCT}{\mathcal{K}}
\newcommand{\Dil}{\mathcal{D}}

\newcommand{\dil}[1]{D^{(#1)}}
\newcommand{\sct}[2]{{K^{(#1)}}_{#2}}

\newcommand{\dilL}[1]{\overleftarrow{D}^{(#1)}}
\newcommand{\sctL}[2]{{\overleftarrow{K}^{(#1)}}_{#2}}

\newcommand{\quasi}{\mathcal{O}}

\newcommand{\arbitrary}[1]{\mathcal{#1}}

\newcommand{\acom}[2]{\bigl\{#1,#2\bigr\}}
\newcommand{\remainder}[1]{\tilde{#1}}

\newcommand{\dfield}{\varphi}

\newcommand{\cutoff}{K}
\newcommand{\ep}{\mathcal{G}}

\newcommand{\Stot}{S}
\newcommand{\Sint}{\mathcal{S}}
\newcommand{\SGauss}{\Stot^{\mathrm{Gauss}}}

\newcommand{\R}{R}
\newcommand{\Ltot}{\hat{L}}

\newcommand{\lrtimes}[3]{{#1} \rtimes {#2}  \ltimes {#3}}
\newcommand{\core}[1]{\mathscr{#1}}

\newcommand{\QEMT}{Q}
\newcommand{\CEMT}{t}
\newcommand{\Bel}{B}
\newcommand{\conserved}[1]{\overline{#1}}
\newcommand{\symmetric}[1]{#1^{\mathrm{sym}}}

\newcommand{\hepth}[1]{hep-th/#1}

\begin{document}

\title{A Wilsonian Energy-Momentum Tensor}
\author{Oliver J.~Rosten}
\email{oliver.rosten@gmail.com}
\affiliation{Unaffiliated \vspace{2ex}}

\begin{abstract}
\vspace{2ex}
	For local conformal field theories, it is shown how to construct an expression for the energy-momentum tensor in terms of a Wilsonian effective Lagrangian. Tracelessness implies a single, unintegrated equation which enforces both the Exact Renormalization Group equation and its partner encoding invariance under special conformal transformations.
\end{abstract}

\maketitle
\tableofcontents

\section{Introduction}

In the context of quantum field theory,
consider the action for a free, scalar field in $\D$-dimensional Euclidean space:
\be
	\SGauss[\dfield] = \hf \Int{x} \partial_\mu \dfield\, \partial_\mu \dfield
.
\label{eq:SGauss}
\ee
What is the scaling dimension of the field? This is, of course, a trivial question to answer: since the theory is non-interacting, we can just use \naive\ power counting to obtain $(\D-2)/2$. However, this method does not carry over to the interacting case; consequently, we seek other, more elaborate ways of answering this question that  are more generalizable.

Our next approach---which, at least in its preliminary incarnation, is in a sense just a somewhat more formal means of power-counting---involves constructing a functional representation of the dilatation generator. First define
\be
	\dil{\delta} = x \cdot \partial + \delta
,
\label{eq:dil^delta}
\ee
where $\delta$ is an \emph{a priori} unknown scaling dimension and, in this context, the dot denotes a contraction of indices. Next introduce
\be
	\Dil = \dil{\delta} \dfield \cdot \fder{}{\dfield}
\ee
where here the dot denotes an integral over the position of the field. Demanding dilatation invariance of the action recovers the previous result:
\be
	 \dil{\delta} \dfield \cdot \fder{}{\dfield} \SGauss[\dfield] = 0
	 \qquad
	 \Rightarrow
	 \qquad
	 \delta = \delta_0 \equiv \frac{\D-2}{2}
.
\ee
As it stands, this method is also of no use in the interacting case, but the approach has an appropriate generalization.

If we wish to retain a representation of the dilatation generator for interacting quantum field theories which is linear in the functional derivative, then we must take it to act on the correlation functions, which are non-local. To preserve a local description, we are forced instead consider representations of the dilatation generator which are (at least) quadratic in derivatives. The Exact Renormalization Group (ERG) provides a particular realization of this:
\be
	\Dil = 
	\dil{\delta} \dfield \cdot \fder{}{\dfield}
	+ \dfield \cdot \ep^{-1} \cdot G \cdot \fder{}{\dfield}
	+ \hf \fder{}{\dfield} \cdot G \cdot \fder{}{\dfield}
,
\label{eq:Dil-ERG}
\ee
where $G$ and $\ep$ incorporate an ultraviolet cutoff function, $\cutoff$, in a manner to be specified later. Scale invariance is enforced by the fixed-point condition:
\be
	\Dil \, e^{-\Stot} = 0
.
\label{eq:ScaleInv}
\ee

A particularly interesting feature of this equation is that, in general, not only is the scaling dimension \emph{a priori} unknown but, so too, is the action. The correct understanding is to recognize the ERG equation as a non-linear eigenvalue equation~\cite{TRM-Elements}. In principle, then, the ERG equation allows the self-consistent determination of the spectrum of local fixed-points. However, the equation itself is fiendishly difficult to attack and, in general, various rather brutal approximation schemes must be employed.

Within the context of approaches that are intrinsically local, are there any options to improve upon the ERG? A clue comes from the fact that~\eq{ScaleInv} is only a statement of scale invariance; it does not automatically incorporate full conformal invariance (for a detailed discussion of the relationship between scale and conformal invariance, see~\cite{Nakayama-ScaleConformal}). Whilst it is true that for many theories of interest scale invariance in fact enhances to conformal invariance, this is not a general property of all solutions of~\eq{ScaleInv}. Indeed, just as~\eq{Dil-ERG} provides a representation of the dilatation generator, so too is there an associated representation of the special conformal generator~\cite{Schafer-Conformal,Representations,HO-Functional}. The form we use is equivalent to that derived in~\cite{Representations}:
\be
	\SCT_\mu
	=
	\sct{\delta}{\mu} \dfield \cdot \fder{}{\dfield}
	+ \dfield \cdot \ep^{-1} \cdot G_\mu \cdot \fder{}{\dfield}
	+ \hf \fder{}{\dfield} \cdot G_\mu \cdot \fder{}{\dfield}
	- \eta \partial_\alpha \dfield \cdot \cutoff^{-1} \cdot G \cdot \fder{}{\dfield}
,
\label{eq:Sct-ERG}
\ee
where $G_\mu$ is related to $G$ (again, with the details deferred until later),
\be
	\sct{\delta}{\mu} = 2x_\mu \bigl(x \cdot \partial + \delta \bigr) - x^2 \partial_\mu
,
\ee
 and $\eta$ is defined via
\be
	\delta = \delta_0 + \eta/2 = \frac{\D-2+\eta}{2}
. 
\ee
 A conformally invariant theory thus satisfies not just~\eq{ScaleInv} but also
\be
	\SCT_\mu \, e^{-\Stot} = 0
.
\label{eq:SCInv}
\ee

However rather than attempting to solve the pair of equations~\eqs{ScaleInv}{SCInv} as they stand, in this paper we shall instead seek a single equation which incorporates full conformal invariance. To understand how such an equation might arise, let us consider a third answer to the question posed at the start of this paper: we will obtain the scaling dimension of our Gaussian theory via consideration of the energy-momentum tensor. This is significantly more involved than either of the above approaches, but it has the merit of providing an interesting and potentially powerful generalization.

For the Gaussian theory, we can take the energy-momentum tensor to be defined by the following three equations:
\begin{subequations}
\begin{align}
	\partial_\alpha T_{\alpha\beta}^{\mathrm{Gauss}} 
	& = 
	-\partial_\beta \dfield \times \fder{\SGauss}{\dfield}
,
\label{eq:divT-G}
\\
	T_{\alpha\beta}^{\mathrm{Gauss}} & = T_{\beta \alpha}^{\mathrm{Gauss}}
,
\label{eq:rotT-G}
\\
	T_{\alpha\alpha}^{\mathrm{Gauss}} & = -\delta \dfield \times \fder{\SGauss}{\dfield}
.
\label{eq:dilT-G}
\end{align}
\end{subequations}
In position space, the $\times$ just represents the product of two quantities at the same location; it can be omitted and is generally used to emphasise the lack of an integral (which would be denoted by a dot).
These three equations encode, respectively, translational, rotational and scale-invariance of the action; however, these three invariances can only be expressed in terms of a single object---the energy momentum tensor---if the theory is in fact conformally invariant.

It is instructive to see how these equations can be used to determine $\delta$. Substituting~\eq{SGauss} into~\eq{divT-G} and rearranging gives:
\be
	\partial_\alpha T_{\alpha\beta}^{\mathrm{Gauss}} 
	=
	\partial_\beta \dfield \times \partial^2 \dfield 
	=
	\partial_\alpha
	\Bigl(
		\partial_\alpha \dfield \times \partial_\beta \dfield 
		- \delta_{\alpha\beta} \smallhf \partial_\lambda \dfield \times \partial_\lambda \dfield
	\Bigr)
.
\ee
Therefore,
\be
	T_{\alpha\beta}^{\mathrm{Gauss}}  = 
	\partial_\alpha \dfield \times \partial_\beta \dfield 
	- \delta_{\alpha\beta} \smallhf \partial_\lambda \dfield \times \partial_\lambda \dfield
	+\partial_\lambda W_{\lambda \alpha \beta},
\ee
where $W_{\lambda \alpha \beta} = -W_{\alpha \lambda \beta}$ vanishes when contracted with $\partial_\alpha$. The condition~\eq{rotT-G} enforces $W_{\lambda \alpha \beta} = W_{\lambda \beta \alpha}$. For the Gaussian theory, where all terms contributing to the energy-momentum tensor must have two derivatives and two powers of the field, it follows that
\be
	\partial_\lambda W_{\lambda \alpha \beta}
	=
	w
	\bigl(
		\delta_{\alpha\beta} \partial^2 -\partial_\alpha \partial_\beta
	\bigr)
	\dfield^2
,
\ee
where $w$ is determined by the condition for scale invariance~\eq{dilT-G}:
\be
	\frac{2-\D}{2} \partial_\alpha \dfield \times \partial_\alpha \dfield
	+ w(\D-1) \partial^2 \dfield^2
	=
	\delta \dfield \times \partial^2 \dfield
.
\ee
A bit of simple algebra reveals that:
\be
	w = \frac{\D-2}{4(\D-1)},
	\qquad
	\delta = 2 w (\D-1) = \frac{\D-2}{2},
\ee
as before.

The generalizations of~\eqss{divT-G}{rotT-G}{dilT-G} appropriate to interacting theories within the framework of the ERG were derived in~\cite{Representations} and subsequently explored in~\cite{Sonoda-EMT}. For a putative fixed-point, it is always possible to construct a symmetric, conserved tensor, as argued in~\cite{Representations} and as we shall explicitly see below. However, dilatation invariance is not automatic and demanding it be satisfied produces a constraint equation, \eq{CERG}, which we shall refer to as the `conformal fixed-point equation'. In contrast to the ERG equation~\eq{ScaleInv} and its partner~\eq{SCInv} this equation is unintegrated---reflecting the fact that it involves the Lagrangian rather than just the action. Moreover, this single equation automatically enforces both~\eqs{ScaleInv}{SCInv}; as such, the set of solutions to the conformal fixed-point equation determines the spectrum of local conformal field theories. Individual solutions self-consistently determine the action and anomalous dimension of the fundamental field, while simultaneously providing the requisite improvement of the energy-momentum tensor such as to render it traceless. It is beyond the scope of this paper to investigate methods of solving the conformal fixed-point equation and thus addressing the question of whether or not it confers an advantage over the plain ERG equation and its special conformal partner.

Before diving into the ERG treatment, in \sect{General} we first recall the generalizations of~\eqss{divT-G}{rotT-G}{dilT-G} in an arbitrary representation of the conformal algebra and then describe a strategy for solving these equations. In \sect{Classical} we apply this method in the context of classical theories, and  illustrate it with some concrete examples. The classical solution forms part of the full ERG solution which is presented in \sect{ERG}, facilitated by some new notation. To clarify the rather technical development, the strategy is summarised in the conclusion with key results rewritten in \app{uncondensed} using standard notation. However, it is worth bearing in mind that, having been obtained, the veracity of the conformal fixed-point equation~\eq{CERG} can be checked very easily.

\section{The Energy-Momentum Tensor}

\subsection{General Considerations}
\label{sec:General}

As emphasised in~\cite{Representations}, if one is to consider various representations of the conformal algebra, then one must be prepared to consider associated representations of the energy-momentum tensor. In this section we work in a local but otherwise arbitrary representation and, to indicate this, use the symbol $\arbitrary{T}_{\alpha\beta}$. As in the introduction, throughout the rest of this paper we work in $\D$-dimensional Euclidean space.

To motivate an appropriate generalization of~\eqss{divT-G}{rotT-G}{dilT-G}, note that the first and last equations both involve the two separate objects, $\dfield$ and $\delta S / \delta \dfield$. The restriction to the Gaussian fixed-point has been lifted since henceforth our interest is in general conformal field theories. Intuitively, the first object, $\dfield$, should have a scaling dimension of $\delta$; from this and a straightforward dimensional analysis, it follows that the second object has scaling dimension $\D-\delta$.

Now, in an arbitrary local representation of the conformal algebra, while we can reasonably expect the quasi-primary field of scaling dimension $\delta$ to be intimately related to $\dfield$, it may have a more complicated form. For example, in an ERG representation, it contains additional terms which depend on the cutoff, but which vanish when the cutoff is removed. Thus, instead of explicitly working with $\dfield$, we prefer the more general $\quasi^{(\delta)}$, which is defined to be a quasi-primary field of scaling dimension $\delta$ \viz\
\be
	\Dil \quasi^{(\delta)} = \dil{\delta}  \quasi^{(\delta)},
	\qquad
	\SCT_\mu \quasi^{(\delta)} = \sct{\delta}{\mu}  \quasi^{(\delta)}
.
\label{eq:Odelta}
\ee
Similarly, we trade $\delta S / \delta \dfield$ for $\quasi^{(\D-\delta)}$, where the latter satsfies
\be
	\Dil \quasi^{(\D-\delta)} = \dil{\D-\delta}  \quasi^{(\D-\delta)},
	\qquad
	\SCT_\mu \quasi^{(\D-\delta)} = \sct{\D-\delta}{\mu}  \quasi^{(\D-\delta)}
.
\label{eq:Od-delta}
\ee
With this in mind, the generalization of~\eqss{divT-G}{rotT-G}{dilT-G} which we seek---and which is fully justified in~\cite{Representations}---reads:
\begin{subequations}
\begin{align}
	\partial_{\alpha} \arbitrary{T}_{\alpha\beta}
	&=
	-\hat{\quasi}^{(\D-\delta)} \times \partial_\beta \quasi^{(\delta)}
,
\label{eq:divT}
\\
	\arbitrary{T}_{\alpha\beta} 
	&= \arbitrary{T}_{\beta\alpha},
\label{eq:symT}
\\
	\arbitrary{T}_{\alpha \alpha}
	& = -\delta \hat{\quasi}^{(\D-\delta)} \times \quasi^{(\delta)}
\label{eq:traceT}
,
\end{align}
\end{subequations}
where
\be
	\bigl[\Dil ,  \hat{\quasi}^{(\D-\delta)} \bigr] = \dil{\delta} \hat{\quasi}^{(\D-\delta)},
	\qquad
	\bigl[\SCT_\mu ,  \hat{\quasi}^{(\D-\delta)} \bigr] = \sct{\delta}{\mu}  \hat{\quasi}^{(\D-\delta)}
.
\label{eq:hatO}
\ee
The origin of the commutator may be seen by acting on~\eq{traceT} with $\Dil$, bearing in mind that $\arbitrary{T}_{\alpha\alpha}$ is expected to be of scaling dimension $\D$. For representations of the conformal algebra linear in functional derivatives, $\hat{\quasi}^{(\D-\delta)}$ just reduces to $\quasi^{(\D-\delta)}$. However, the ERG representation---which is of particular interest to us---is quadratic in derivatives and in this case it turns out that $\hat{\quasi}^{(\D-\delta)}$ can be constructed from ${\quasi}^{(\D-\delta)}$ by adding a functional derivative term, as we will see later.

A theory is conformal if and only if a solution exists to~\eqss{divT}{symT}{traceT}. However, for theories which are scale invariant but not fully conformal, while a solution may be constructed for~\eqs{divT}{symT}, the trace of this object violates~\eq{traceT}. Either way, acting on~\eq{divT} with $\Dil$, and using~\eqss{Odelta}{Od-delta}{hatO}, it follows that $\partial_\alpha \arbitrary{T}_{\alpha\beta}$ has scaling dimension $\D+1$. This implies that $\arbitrary{T}_{\alpha\beta}$ has, as expected, scaling dimension $\D$ but only up to transverse terms. For fully conformal theories, \eq{traceT} guarantees that any such transverse terms do not prevent
$\arbitrary{T}_{\alpha\beta}$ from being a \emph{bona-fide} scaling field of dimension $\D$. On the other hand, if the theory is merely scale invariant, it may be that the $\arbitrary{T}_{\alpha\beta}$ constructed as a solution to~\eqs{divT}{symT} is not a scaling field. Nevertheless, in this case, $\arbitrary{T}_{\alpha\beta}$ may be redefined so as to cancel this transverse term, since \eqs{divT}{symT} are invariant under transverse modifications.

The remainder of this section is devoted to understanding the structure of solutions to~\eqss{divT}{symT}{traceT}.
Suppose that there exists a recipe to extract a conserved (though not necessarily symmetric) contribution to the energy-momentum tensor, $\conserved{\mathcal{T}}_{\alpha\beta}$. In this case, the  solution to~\eq{divT} can, along the lines of the Belinfante tensor, be expressed as:
\be
	\arbitrary{T}_{\alpha\beta}
	=
	\overline{\arbitrary{T}}_{\alpha\beta} + \partial_\lambda \arbitrary{\Bel}_{\lambda\alpha\beta},
\label{eq:Tbar}
\ee
where
\be
	\arbitrary{\Bel}_{\lambda\alpha\beta} = - \arbitrary{\Bel}_{\alpha\lambda\beta}
.
\label{eq:B-asym}
\ee
Integrating~\eq{Tbar} and using~\eq{symT} it is apparent that, for local solutions, $\conserved{\arbitrary{T}}_{\alpha\beta}$ is symmetric up to total derivatives.  Therefore, for some 
$\arbitrary{F}_{\lambda \alpha \beta}$ and some
symmetric $\conserved{\arbitrary{T}}^{\mathrm{sym}}_{\alpha\beta}$ we may write
\be
	\conserved{\arbitrary{T}}_{\alpha\beta} 
	= \conserved{\arbitrary{T}}^{\mathrm{sym}}_{\alpha\beta}
	+ \partial_\lambda \arbitrary{F}_{\lambda \alpha \beta}
,
\label{eq:TbarSym}
\ee
whereupon it follows that
\be
	\overline{\arbitrary{T}}_{\alpha\beta}
	-
	\overline{\arbitrary{T}}_{\beta \alpha}
	=
	\partial_\lambda
	\bigl(
		\arbitrary{F}_{\lambda \alpha \beta}
		-
		\arbitrary{F}_{\lambda \beta \alpha }
	\bigr)
.
\label{eq:Tbar-asym}
\ee
Combining with~\eqs{symT}{Tbar} yields
\be
	\partial_\lambda
	\bigl(
		\arbitrary{F}_{\lambda \alpha \beta}
		-
		\arbitrary{F}_{\lambda \beta \alpha }
	\bigr)
	=
	-\partial_\lambda
	\bigl(
		\arbitrary{B}_{\lambda \alpha \beta}
		-
		\arbitrary{B}_{\lambda \beta \alpha }
	\bigr)
\label{eq:BandF}
\ee

The goal now is to solve the conditions~\eqs{B-asym}{BandF} for $\arbitrary{B}_{\lambda\alpha\beta}$ in terms of $\arbitrary{F}_{\lambda\alpha\beta}$. To this end, it is useful to follow Belinfante and Rosenfeld~\cite{Belinfante, Rosenfeld}. Defining
\be
	\tau_{\lambda \alpha \beta}
	\equiv
	\arbitrary{F}_{\lambda \alpha \beta}
	-
	\arbitrary{F}_{\lambda \beta \alpha }
,
\label{eq:tau}
\ee
the solution is given by
\be
	\arbitrary{B}_{\lambda \alpha \beta}
	=
	\hf
	\bigl(
		\tau_{\beta\lambda\alpha}
		- \tau_{\lambda \alpha \beta}
		- \tau_{\alpha \beta \lambda}
	\bigr)
	+
	\partial_\mu 
    \arbitrary{Y}_{\alpha \lambda \beta \mu}
\label{eq:B}
,
\ee
where $\arbitrary{Y}_{\alpha \lambda \beta \mu}$ has the following symmetries:
\be
	\arbitrary{Y}_{\alpha \lambda \beta \mu}
	=
	\arbitrary{Y}_{\beta \mu \alpha \lambda }
	=
	-\arbitrary{Y}_{ \lambda  \alpha \beta \mu}
.
\label{eq:Y-syms}
\ee
Substituting~\eq{tau} into~\eq{B}, it follows immediately that the desired constraints~\eqs{B-asym}{BandF} are satisfied. Therefore, substituting~\eq{B} into~\eq{Tbar} yields a tensor which is both conserved and symmetric:
\be
	\arbitrary{T}_{\alpha\beta}
	=
	\conserved{\arbitrary{T}}_{\alpha\beta}
	+
	\hf \partial_\lambda
	\bigl(
		\arbitrary{F}_{\alpha \lambda \beta} 
		+ \arbitrary{F}_{\beta \lambda \alpha} 
		+ \arbitrary{F}_{\lambda \beta \alpha}
		- \arbitrary{F}_{\lambda \alpha \beta} 
		- \arbitrary{F}_{\beta \alpha \lambda} 
		-  \arbitrary{F}_{\alpha \beta \lambda}
	\bigr)
	+
	\partial_\lambda
	\arbitrary{W}_{\lambda \alpha \beta} 
\label{eq:symmetricEMT}
\ee
where, for convenience, we have defined
\be
	\arbitrary{W}_{\lambda \alpha \beta} 
	\equiv
	\partial_\mu 
	 \arbitrary{Y}_{\alpha \lambda \beta \mu}
\ee
which inherits from~\eq{Y-syms} the following:
\be
	\partial_\lambda \arbitrary{W}_{\lambda \beta \alpha} 
	= \partial_\lambda  \arbitrary{W}_{\lambda \alpha \beta }
,
	\qquad   
	\arbitrary{W}_{\lambda \alpha \beta}
	= - \arbitrary{W}_{\alpha \lambda \beta}
.
\label{eq:W-conditions}
\ee

We now turn our attention to the trace of the energy-momentum tensor, starting with a return to general properties of~\eqss{divT}{symT}{traceT}. While a solution to~\eqs{divT}{symT} does not imply a solution to~\eq{traceT}, under the assumption of dilatation invariance it does imply that~\eq{traceT} holds in integrated form. This may be argued by multiplying~\eq{divT} by $x_\beta$ and then integrating over all space. Integrating by parts on the \lhs\ and massaging the \rhs\ gives
\be
	\Int{x} \arbitrary{T}_{\alpha \alpha}(x)
	=
	\Int{x} \hat{\quasi}^{(\D-\delta)}(x) \bigl(x\cdot \partial + \delta \big) \quasi^{(\delta)}(x)
	-
	\delta 
	\Int{x} \hat{\quasi}^{(\D-\delta)}(x) \quasi^{(\delta)}(x)
.
\ee
The first term on the \rhs\ is just an expression of dilatation invariance. This is readily seen for a classical representation, where it reduces to $\dil{\delta} \dfield \cdot \delta S / \delta \dfield$. That it holds more generally is discussed fully in~\cite{Representations}. The point is that this term vanishes, revealing an integrated form of~\eq{traceT}.

Therefore, given an assumption of dilatation invariance, a solution to~\eqs{divT}{symT} solves~\eq{traceT} up to a total derivative. Temporarily supposing that $\arbitrary{W}_{\lambda \alpha \beta} = 0$, this implies that, for some $\mathcal{H}_\lambda$,
\[
	\arbitrary{T}_{\alpha \alpha}
	= -\delta \hat{\quasi}^{(\D-\delta)} \times \quasi^{(\delta)} - \partial_\lambda \mathcal{H}_\lambda
.
\]
Inspecting our explicit solution~\eq{symmetricEMT}, the question as to whether the energy-momentum tensor can be improved amounts to asking whether it is possible to find a suitable $\arbitrary{W}_{\lambda \alpha \beta}$ to remove the unwanted total derivative term~\cite{Representations}. Therefore, we seek a solution to~\eqs{W-conditions}{W-extra} together with
\be
	\partial_\lambda \arbitrary{W}_{\lambda \alpha \alpha} = \partial_\lambda \arbitrary{H}_\lambda
.
\label{eq:W-extra}
\ee
This may be accomplished by introducing $\arbitrary{H}_{\tau\lambda}$ and taking
\be
	\arbitrary{H}_{\lambda} = \partial_\tau \arbitrary{H}_{\tau\lambda}
\ee
in terms of which we recover Polchinski's solution~\cite{Pol-ScaleConformal}%
\footnote{For non-unitary theories in $\D>3$, it is conceivable that there is an ambiguity in the energy-momentum tensor. This arises since a conformal primary may exist from which a conserved, symmetric, identically traceless tensor can be constructed~\cite{HO-EMT,Representations}. Such an ambiguity will not be explicitly treated in this paper.}%
:
\begin{subequations}
\begin{align}
\nonumber
	\partial_\lambda \arbitrary{W}_{\lambda \alpha \beta}
	& =
	\frac{1}{2-\D}
	\bigl(
		\partial_\alpha \partial_\tau \arbitrary{H}_{\tau \beta} +
		\partial_\beta \partial_\tau \arbitrary{H}_{\tau \alpha}
		-\partial^2 \arbitrary{H}_{\alpha \beta}
		-\delta_{\alpha \beta} \partial_\tau \partial_\lambda \arbitrary{H}_{\tau \lambda} 
	\bigr)
\\
	&
	\qquad+
	\frac{1}{(2-\D)(\D-1)}
	\bigl(
		\delta_{\alpha \beta} \partial^2
		-\partial_{\alpha} \partial_\beta
	\bigr) \arbitrary{H}_{\tau \tau}
	&
	\mathrm{for}\ \D > 2
,
\label{eq:Pol-condition-d>2}
\\
	\partial_\lambda \arbitrary{W}_{\lambda \alpha \beta}
	& =
	\frac{1}{1-\D}
	\bigl(
		\partial_{\alpha} \partial_\beta
		-\delta_{\alpha \beta} \partial^2
	\bigr)  \arbitrary{H}
	&
	\mathrm{for}\ \D = 2	
,
\label{eq:Pol-condition-d=2}
\end{align}
\end{subequations}
where, in $\D=2$, $\arbitrary{H}_{\tau \lambda} = \delta_{\tau\lambda} \arbitrary{H}$.  Notice that, for $\D>2$, it may be that  $\arbitrary{H}_{\tau \lambda}$ is determined only up to a transverse piece: if a quasi-local $f$ exists of scaling dimension $\D-2$ then, given a solution $\arbitrary{H}_{\tau \lambda}$, we may generate a one-parameter family of  solutions
$\arbitrary{H}_{\tau \lambda}(a) = \arbitrary{H}_{\tau \lambda} + 
a (\delta_{\tau\lambda} \partial^2 - \partial_\tau \partial_\lambda) f$. However, the \rhs\ of~\eq{Pol-condition-d>2} is readily seen to be independent of $a$ and so this ambiguity in $\arbitrary{H}_{\tau \lambda}$ has no effect on the energy-momentum tensor.

Having discussed these generalities, let us now return to our explicit solution~\eq{symmetricEMT} and take the trace:
\be
	\arbitrary{T}_{\alpha\alpha} = \conserved{\arbitrary{T}}_{\alpha\alpha}
	+
	\partial_\lambda
	\bigl(
		\arbitrary{F}_{\alpha \lambda \alpha} 
		- \arbitrary{F}_{\alpha \alpha \lambda} 
		+ \arbitrary{W}_{\lambda \alpha \alpha}
	\bigr)
.
\label{eq:traceOfEMT}
\ee
Comparing this with~\eq{traceT} gives a consistency condition:
\be
	 \conserved{\arbitrary{T}}_{\alpha\alpha}
	+
	\partial_\lambda
	\bigl(
		\arbitrary{F}_{\alpha \lambda \alpha} 
		- \arbitrary{F}_{\alpha \alpha \lambda} 
		+ \arbitrary{W}_{\lambda \alpha \alpha}
	\bigr)
	=
	-\delta \hat{\quasi}^{(\D-\delta)} \times \quasi^{(\delta)}
\label{eq:GenDilCondition}
\ee
which, if satisfied, amounts to the energy-momentum tensor being improvable, so that it is not only conserved and symmetric, but also traceless. Substituting~\eq{W-extra} into~\eq{GenDilCondition}
yields
\be
	-\partial_\lambda \partial_\tau \arbitrary{H}_{\tau \lambda}
	=
 	\conserved{\arbitrary{T}}_{\alpha\alpha}
	+
	\partial_\lambda
	\bigl(
		\arbitrary{F}_{\alpha \lambda \alpha} 
		- \arbitrary{F}_{\alpha \alpha \lambda} 
	\bigr)
	+\delta \hat{\quasi}^{(\D-\delta)} \times \quasi^{(\delta)}
,
\label{eq:ConstraintGeneral}
\ee
which will be central to our analysis below and is the basis for the conformal fixed-point equation.
Note that any terms containing two or more total derivatives on the \rhs\ can be absorbed by redefining $\arbitrary{H}_{\tau\lambda}$; this will be exploited in subsequent sections.

An important subtlety is that, for~\eq{ConstraintGeneral} to be soluble, it seems there must be an interesting conspiracy, since the \rhs\ is not manifestly $\order{\partial^2}$. The resolution is as follows: integrating this equation yields
\be
	\Int{x} \conserved{\arbitrary{T}}_{\alpha\alpha}(x)
	+ \delta  \hat{\quasi}^{(\D-\delta)} \cdot \quasi^{(\delta)}
	= 0,
\label{eq:GeneralDil}
\ee
whereas first multiplying by $x_\mu$ gives
\be
	\Int{x}
	\biggl(
		x_\mu \conserved{\arbitrary{T}}_{\alpha\alpha}(x)
		-
			\arbitrary{F}_{\alpha \mu \alpha} (x)
			+ \arbitrary{F}_{\alpha \alpha \mu} (x)
		+
		\delta\,
		\hat{\quasi}^{(\D-\delta)}(x)  x_\mu
	    \quasi^{(\delta)}(x)
	\biggr)	   
	   	= 0
.
\label{eq:GeneralSCT}
\ee
Therefore, existence of a solution to~\eq{ConstraintGeneral} implies one scalar and one vector condition on the action, both independent of $\arbitrary{H}_{\tau\lambda}$; respectively, these are naturally associated with dilatation and special conformal invariance.

There is an additional, interesting subtlety in $\D=2$: it is possible to have a theory for which the action satisfies both dilatation and special conformal invariance but, nevertheless, the quantum theory is not conformal! This can arise if it is not possible to express $\arbitrary{H}_{\tau \lambda} = \delta_{\tau \lambda} \arbitrary{H}$; we shall encounter this later when examining the higher derivative $\dfield \, \partial^4 \dfield$ theory, which suffers from particularly bad infrared behaviour in $\D=2$.

\subsection{Classical Theories}
\label{sec:Classical}

\subsubsection{Analysis}

In this section we apply the general methodology of \sect{General} in a classical context. On the one hand, this will provide some experience with the advocated approach; on the other the results of this section will form part of the full quantum field theoretic result. Classically, the appropriate form for the defining equations of the energy-momentum tensor, \eqss{divT}{symT}{traceT} are:
\begin{subequations}
\begin{align}
	\partial_\alpha \CEMT_{\alpha\beta}
	& = 
	-\partial_\beta \dfield \times \fder{\Stot}{\dfield}
,
\label{eq:divT-C}
\\
	\CEMT_{\alpha\beta} & = \CEMT_{\beta \alpha}
,
\label{eq:rotT-C}
\\
	\CEMT_{\alpha\alpha} & = -\delta \dfield \times \fder{\Stot}{\dfield}
,
\label{eq:dilT-C}
\end{align}
\end{subequations}
where $\CEMT_{\alpha\beta}$ denotes the classical energy-momentum tensor. Note that these equations are exactly of the form as in the introduction, \eqss{divT-G}{rotT-G}{dilT-G} and, as with the earlier equations, respectively encode translational, rotational and scale-invariance. Throughout this section, when we talk of an energy-momentum tensor which is conserved/traceless, we mean that it is conserved/traceless up to terms which vanish on the equations of motion.

To proceed,  let us take $\Ltot$ to be an arbitrary element of the equivalence class of objects that integrate to the Wilsonian effective action, \viz
\be
	\Int{x} \Ltot(x) = \Stot[\dfield]
,
\label{eq:Lhat}
\ee
but for which any total-derivative contributions have been discarded.
For example, the Gaussian theory has $\{\Ltot\} = \{\frac{1-a}{2}\partial_\mu\dfield \partial_\mu \dfield - \frac{a}{2} \dfield \partial^2 \dfield; \ -\infty < a < \infty\}$ with $a=0$ singled out as \emph{the} Lagrangian.
We henceforth demand quasi-locality, meaning that we restrict our attention to Lagrangians (and hence actions) which exhibit a derivative expansion, \viz
\be
	\Ltot(x) = V(\dfield) + Z(\dfield) \partial_\mu \dfield\, \partial_\mu \dfield + \ldots,
\ee
where $V$ and $Z$ do not contain any derivatives, and the ellipsis denotes terms higher order in derivatives. 

To aid the analysis, define:
\be
	\partialprod{\alpha}{j}{i}
	\equiv
	\Biggl\{
	\begin{array}{cl}
		\prod_{k=j}^i \partial_{\alpha_k} & i \geq j,
	\\
		1 & i < j.
	\end{array}
\label{eq:condensed_partials}
\ee
Using this notation, we have
\be
	\fder{\Stot}{\dfield}
	=
	\sum_{i=0}^\infty (-1)^i
	\partialprod{\sigma}{1}{i}
	\pder{\Ltot}{(\partialprod{\sigma}{1}{i} \dfield)}
,
\label{eq:dS/dphi}
\ee
where it is understood that after expanding out using~\eq{condensed_partials}, the repeated indices $\sigma_1, \ldots \sigma_n$ are summed over. Defining
\be
	S_\alpha
	=
	-\sum_{i=0}^\infty
	(-1)^i
	\partialprod{\sigma}{1}{i}
	\pder{\Ltot}{(\partial_\alpha \partialprod{\sigma}{1}{i} \dfield)}
,
\label{eq:S_a}
\ee
it is apparent that
\be
	\partial_\alpha S_\alpha
	=
	\fder{\Stot}{\dfield}
	-
	\pder{\Ltot}{\dfield}
.
\ee

Courtesy of the chain rule,
\be
	\partial_\alpha \Ltot
	=
	\sum_{i=0}^\infty
	\partial_\alpha \partialprod{\sigma}{1}{i} \dfield
	\pder{\Ltot}{(\partialprod{\sigma}{1}{i} \dfield)}
.
\label{eq:chain}
\ee
Turning our attention to the conservation equation~\eq{divT-C} the above results imply that
\begin{multline}
	\partial_\beta \dfield \times \fder{\Stot}{\dfield}
	=
	\partial_\alpha
	\bigl(
		\partial_\beta \dfield \times S_\alpha + \delta_{\alpha \beta} \Ltot
	\bigr)
\\
	+
	\partial_\alpha \partial_\beta \dfield
	\sum_{i=0}^\infty
	(-1)^i
	\partialprod{\sigma}{1}{i}
	\pder{\Ltot}{(\partial_\alpha \partialprod{\sigma}{1}{i} \dfield)}
	-
	\sum_{i=1}^\infty	
	\partial_\beta 
	\partialprod{\sigma}{1}{i} \dfield
	\pder{\Ltot}{(\partialprod{\sigma}{1}{i} \dfield)}
.
\label{eq:NotYetTotalDeriv}
\end{multline}
In order to generate a contribution to the energy-momentum tensor, we must convert the whole of the \rhs\ into a total derivative. As preparation for manipulating the middle term, 
observe that for some $A$, $B$ and $j<i$
\be
	\partialprod{\sigma}{1}{j} A \times \partialprod{\sigma}{j+1}{i} B
	=
	\partial_{\sigma_{j+1}}
	\bigl(
		\partialprod{\sigma}{1}{j} A \times \partialprod{\sigma}{j+2}{i} B
	\bigr)
	-
	\partialprod{\sigma}{1}{j+1} A \times \partialprod{\sigma}{j+2}{i} B
.
\ee
Feeding this result back into the final term and iterating yields, for $j<i$:
\be
	\partialprod{\sigma}{1}{j} A \times \partialprod{\sigma}{j+1}{i} B
	=
	(-1)^{i+j} \partialprod{\sigma}{1}{i} A \times B
	-\sum_{k=j+1}^i (-1)^{j+k}
	\partial_{\sigma_{k}}
	\bigl(
		\partialprod{\sigma}{1}{k-1} A \times \partialprod{\sigma}{k+1}{i} B
	\bigr)
.
\label{eq:DAxDB}
\ee
We now wish to apply this result, with $j=0$, to the middle term of~\eq{NotYetTotalDeriv}. Since this
may only be done for $i\geq 1$, we first separate off the $i=0$ term before proceeding. This yields:
\begin{multline}
	\partial_\alpha \partial_\beta \dfield
	\sum_{i=0}^\infty
	(-1)^i
	\partialprod{\sigma}{1}{i}
	\pder{\Ltot}{(\partial_\alpha \partialprod{\sigma}{1}{i} \dfield)}
	=
	\sum_{i=0}^\infty
	\partial_\beta \partial_\alpha \partialprod{\sigma}{1}{i} \dfield
	\times
	\pder{\Ltot}{(\partial_\alpha \partialprod{\sigma}{1}{i}\dfield)}
\\
	-
	\sum_{i=1}^\infty \sum_{j=1}^i (-1)^{i+j}
	\partial_{\sigma_j}
	\biggl(
		\partialprod{\sigma}{1}{j-1} \partial_\alpha \partial_\beta \dfield
		\times \partialprod{\sigma}{j+1}{i}
		 \pder{\Ltot}{(\partial_\alpha \partialprod{\sigma}{1}{i} \dfield)}
	\biggr)
.
\end{multline}
The first term cancels the final one of~\eq{NotYetTotalDeriv} leaving, after relabelling the dummy indices
$\alpha \leftrightarrow \sigma_j$,
\begin{multline}
	\partial_\beta \dfield \times \fder{\Stot}{\dfield}
	=
	\partial_\alpha
	\biggl(
		\partial_\beta \dfield \times S_\alpha + \delta_{\alpha \beta} \Ltot
		-
		\sum_{i=1}^\infty \sum_{j=1}^i (-1)^{i+j}
		\partialprod{\sigma}{1}{j} \partial_\beta \dfield
		\times \partialprod{\sigma}{j+1}{i}
		 \pder{\Ltot}{(\partial_\alpha \partialprod{\sigma}{1}{i}\dfield)}
	\biggr)
.
\end{multline}
This equation is s a consequence of translation invariance.
Using~\eq{S_a} to substitute for $S_\alpha$, the resulting term can be absorbed into the final one above by replacing both lower limits of the sums over $j$ and $i$ with zero. Recalling~\eq{Tbar},
we deduce the following contribution to the energy-momentum tensor:
\be
	\conserved{\CEMT}_{\alpha\beta} =
	-\delta_{\alpha\beta} \Ltot
	+
	\sum_{i=0}^\infty
	\sum_{j=0}^i (-1)^{i+j}
	\partialprod{\sigma}{1}{j} \partial_\beta \dfield
	\times
	\partialprod{\sigma}{j+1}{i}
	\pder{\Ltot}{(\partial_\alpha \partialprod{\sigma}{1}{i} \dfield)}
.
\label{eq:tbar_ab}
\ee
To extract $f_{\lambda \alpha \beta}$ (\ie\ the classical version of $\arbitrary{F}_{\lambda \alpha \beta}$) we need to split $\conserved{\CEMT}_{\alpha\beta}$ into a symmetric piece plus a total derivative, as in~\eq{TbarSym}. Utilizing~\eq{DAxDB} gives:
\begin{multline}
	\conserved{\CEMT}_{\alpha\beta} =
	-\delta_{\alpha\beta} \Ltot
	+
	\sum_{i=0}^\infty (i+1)
	\partialprod{\sigma}{1}{i} \partial_\beta \dfield 
	\times \pder{\Ltot}{(\partialprod{\sigma}{1}{i} \partial_\alpha \dfield)}
\\
	-
	\partial_\lambda
	\sum_{i=1}^\infty
	\sum_{j=0}^{i-1}
	\sum_{k=j+1}^i
	(-1)^{i+k}
	\delta_{\lambda \sigma_k}
	\partialprod{\sigma}{1}{k-1}
	\partial_\beta \dfield
	\times
	\partialprod{\sigma}{k+1}{i}
	\pder{\Ltot}{(\partial_\alpha\partialprod{\sigma}{1}{i} \dfield)}
.
\label{eq:Classical-Conserved}
\end{multline}
The key point is that the first two terms are both symmetric under $\alpha \leftrightarrow \beta$. This is manifest for $\delta_{\alpha\beta} \Ltot$. For the second term, we argue as follows.
First, observe that 
\be
	\sum_{i=0}^\infty (i+1)
	\partialprod{\sigma}{1}{i} \partial_\beta \dfield 
	\times \pder{\Ltot}{(\partialprod{\sigma}{1}{i} \partial_\alpha \dfield)}
	=
	\sum_{i=1}^\infty
	\bigl[ \partialprod{\sigma}{1}{i}, x_\alpha \partial_\beta \bigr] \dfield
	\times
	\pder{\Ltot}{(\partialprod{\sigma}{1}{i}\dfield)}
.
\ee
With this in mind, we utilize rotational invariance of the action, together with~\eq{dS/dphi} and integration by parts:
\begin{align}
&
	\Int{x} \bigl(x_\alpha \partial_\beta - x_\beta \partial_\alpha \bigr) \dfield \,
	\fder{\Stot}{\dfield}
	=0
\nonumber
\\
	\Rightarrow
&
	\Int{x} \sum_{i=0}^\infty \bigl(x_\alpha \partial_\beta - x_\beta \partial_\alpha \bigr) 
	\partialprod{\sigma}{1}{i} \dfield
	\pder{\Ltot}{(\partialprod{\sigma}{1}{i} \dfield)}	
	+
	\Int{x} 
	\sum_{i=1}^\infty
	\bigl[ \partialprod{\sigma}{1}{i}, x_\alpha \partial_\beta - x_\beta \partial_\alpha \bigr]
	\dfield
	\pder{\Ltot}{(\partialprod{\sigma}{1}{i} \dfield)}
	= 0
.
\end{align}
The first term vanishes after using the chain rule and integrating by parts and so we conclude that rotational invariance alone is sufficient to ensure that the integrand of the final piece vanishes, at least up to total derivative terms. However, we can go further by exploiting quasi-locality of $\Ltot$. Recall that this is a statement that $\Ltot$ has a derivative expansion; of course, since $\Ltot$ is a scalar, all partial derivatives must be paired up. In the integrand under analysis, the effect on $\Ltot$ of the $[\partialprod{\sigma}{1}{i} , x_\alpha \partial_\beta]$ term is to split all such pairs up into $\partial_\alpha$ and $\partial_\beta$. Since all possible splittings are summed over, the result is symmetric under interchange of indices and so we conclude that
\be
	\sum_{i=1}^\infty
	\bigl[ \partialprod{\sigma}{1}{i}, x_\alpha \partial_\beta - x_\beta \partial_\alpha \bigr]
	\dfield
	\pder{\Ltot}{(\partialprod{\sigma}{1}{i} \dfield)}
	= 0
.
\label{eq:L-RotSymmetry}
\ee

Returning to~\eq{Classical-Conserved}, we use the
recipe~\eq{TbarSym} to extract
\begin{subequations}
\begin{align}
	\symmetric{\conserved{\CEMT}_{\alpha\beta}} 
	& = 
	-\delta_{\alpha\beta} \Ltot
	+
	\sum_{i=1}^\infty
	\bigl[\partialprod{\sigma}{1}{i}, x_\alpha\partial_\beta \bigr] \dfield 
	\times \pder{\Ltot}{(\partialprod{\sigma}{1}{i}  \dfield)}
\label{eq:t_ab}
,
\\
	f_{\lambda \alpha \beta} 
	&=
	-
	\sum_{i=1}^\infty
	\sum_{k=1}^i
	k (-1)^{i+k}
	\delta_{\lambda \sigma_k}
	\partialprod{\sigma}{1}{k-1}
	\partial_\beta \dfield
	\times
	\partialprod{\sigma}{k+1}{i}
	\pder{\Ltot}{(\partial_\alpha\partialprod{\sigma}{1}{i} \dfield)}
\label{eq:f_lab}
\end{align}
\end{subequations}
where, in the final expression, the multiple summations have
been simplified. From~\eqs{t_ab}{f_lab} we can construct the full energy-momentum tensor according to~\eq{symmetricEMT}, at least up to the conserved, symmetric $\partial_\lambda w_{\lambda \alpha \beta}$. To determine the latter, we take the trace---for which we can use~\eq{traceOfEMT}---and compare with~\eq{dilT-C}. Mimicking \sect{General}, we take $\partial_\lambda w_{\lambda \alpha \alpha} = \partial_\lambda \partial_\tau h_{\tau\lambda}$ yielding:
\begin{multline}
	\CEMT_{\alpha \alpha}
	=  \partial_\lambda \partial_\tau h_{\tau\lambda}
	-d \Ltot
	+
	\sum_{i=1}^\infty
	\bigl[ \partialprod{\sigma}{1}{i}, x\cdot\partial \bigr] \dfield
	\times
	\pder{\Ltot}{(\partialprod{\sigma}{1}{i}\dfield)}
\\
	+
	\bigl(
		\delta_{\omega \lambda} \delta_{\rho\sigma}
		- 2\delta_{\omega \rho} \delta_{\sigma\lambda}
	\bigr)
	\partial_\lambda
	\sum_{i=1}^\infty
	\sum_{k=1}^i
	k
	(-1)^{i+k}
	\partialprod{\sigma}{1}{k-1}
	\partial_\omega \dfield \times
	\partialprod{\sigma}{k+1}{i}
	\pder{\Ltot}{(\partial_\rho \partial_\sigma \partialprod{\sigma}{1}{k-1}\partialprod{\sigma}{k+1}{i} 
	\dfield)}
.
\label{eq:t_aa}
\end{multline}
Rather than jumping straight to~\eq{ConstraintGeneral}, we 
recall the comments under~\eqs{Pol-condition-d>2}{Pol-condition-d=2}: the game now is to simplify this expression by absorbing various $\order{\partial^2}$ terms into the first term on the \rhs.
We begin by noting that:
\be
	\sum_{i=1}^\infty
	\sum_{k=1}^i
	k
	\partialprod{\sigma}{1}{k-1}
	\partialprod{\sigma}{k+1}{i}
	\partial_\omega \dfield \times
	\pder{\Ltot}{(\partial_\rho \partial_\sigma \partialprod{\sigma}{1}{k-1}\partialprod{\sigma}{k+1}{i} 
	\dfield)}
	=
	\hf
	\sum_{i=2}^\infty
	\bigl[
		\bigl[
			\partialprod{\sigma}{1}{i}, x_\sigma
		\bigr],
		x_\rho \partial_\omega
	\bigr]
	\dfield
	\times
	\pder{\Ltot}{(\partialprod{\sigma}{1}{i} \dfield)}
.
\ee
To see this, start by shifting $i \rightarrow i-1$ on the \lhs, so that the sum over $i$ now starts from 2. Permutation of the dummy indices on the \lhs\ means that each term arising from the sum over $k$ (which now runs to $i-1$) is identical and therefore there are $i(i-1)/2$ such terms. It is easy to see that this matches the \rhs. To exploit this result in~\eq{t_aa} we must first re-express the final term using a variant of~\eq{DAxDB}:
\be
	\partialprod{\sigma}{1}{k-1} A
	\times
	\partialprod{\sigma}{k+1}{i} B
	=
	(-1)^{i+k}
	\partialprod{\sigma}{1}{k-1} 
	\partialprod{\sigma}{k+1}{i} A \times B
	-\sum_{l=k+1}^i (-1)^{k+l}
	\partial_{\sigma_l}
	\bigl(
		\partialprod{\sigma}{1}{k-1} 
	    \partialprod{\sigma}{k+1}{l-1} A 
	    \times 
	    \partialprod{\sigma}{l+1}{i} B
	\bigr)
,
\ee
whereupon we obtain
\begin{multline}
	\CEMT_{\alpha\alpha} 
	= 
	\partial_\lambda \partial_\tau \remainder{h}_{\tau \lambda}
	-d \Ltot
	+
	\sum_{i=1}^\infty
	\bigl[ \partialprod{\sigma}{1}{i}, x\cdot\partial \bigr] \dfield
	\times
	\pder{\Ltot}{(\partialprod{\sigma}{1}{i}\dfield)}
\\
	+
	\hf
	\bigl(
		\delta_{\omega \lambda} \delta_{\rho\sigma}
		- 2\delta_{\omega \rho} \delta_{\sigma\lambda}
	\bigr)
	\partial_\lambda
	\sum_{i=2}^\infty
	\bigl[
		\bigl[
			\partialprod{\sigma}{1}{i}, x_\sigma
		\bigr],
		x_\rho \partial_\omega
	\bigr]
	\dfield
	\times
	\pder{\Ltot}{(\partialprod{\sigma}{1}{i} \dfield)}
,
\label{eq:taa}
\end{multline}
with
\begin{multline}
	h_{\tau \lambda}
	=
		 \remainder{h}_{\tau \lambda}
		+
		\bigl(
			\delta_{\omega \lambda} \delta_{\rho\sigma}
			- 2\delta_{\omega \rho} \delta_{\sigma\lambda}
		\bigr)
		\sum_{i=2}^\infty
		\sum_{k=1}^{i-1}
		\sum_{l=k+1}^i
		k
		(-1)^{i+l}
		\partialprod{\sigma}{1}{k-1}
		\partialprod{\sigma}{k+1}{l-1}
		\partial_\omega \dfield 
\\
		\times
		\partialprod{\sigma}{l+1}{i}
		\pder{\Ltot}{
			(\partial_\rho \partial_\sigma \partial_\tau 
			\partialprod{\sigma}{1}{k-1}\partialprod{\sigma}{k+1}{l-1} 
			\partialprod{\sigma}{l+1}{i}\dfield)
		}
.
\label{eq:h'}
\end{multline}
In accord with the discussion under~\eq{ConstraintGeneral} we have ignored any additional transverse contributions.
Comparing~\eq{taa} with~\eq{dilT-C} we arrive at the following constraint for CFTs:
\begin{multline}
	- \partial_\lambda \partial_\tau \remainder{h}_{\tau \lambda}
	=
	-\D\Ltot
	+
	\sum_{i=1}^\infty
	\bigl[ \partialprod{\sigma}{1}{i}, x\cdot\partial \bigr] \dfield
	\times
	\pder{\Ltot}{(\partialprod{\sigma}{1}{i}\dfield)}
	+ \delta \dfield \times \fder{\Stot}{\dfield}
\\
	+
	\hf
	\bigl(
		\delta_{\omega \lambda} \delta_{\rho\sigma}
		- 2\delta_{\omega \rho} \delta_{\sigma\lambda}
	\bigr)
	\partial_\lambda
	\sum_{i=2}^\infty
	\bigl[
		\bigl[
			\partialprod{\sigma}{1}{i}, x_\sigma
		\bigr],
		x_\rho \partial_\omega
	\bigr]
	\dfield
	\times
	\pder{\Ltot}{(\partialprod{\sigma}{1}{i} \dfield)}
.
\label{eq:ClassicalConstraint}
\end{multline}
As mentioned at the end of \sect{General}, the existence of solutions to this equation implies two conditions that must be satisfied by the action. This will allow us to check the consistency of~\eq{ClassicalConstraint}.
First of all, we integrate it directly. The total derivative terms vanish. To process the surviving terms, observe that
\begin{align}
	\Int{x} x \cdot \partial \dfield \, \fder{\Stot}{\dfield}
&	=
	\Int{x} x \cdot \partial \dfield
	\sum_{i=0}^\infty (-1)^i
	\partialprod{\sigma}{1}{i}
	\pder{\Ltot}{(\partialprod{\sigma}{1}{i} \dfield)}
\nonumber
\\
&
	=
	\Int{x}
	\biggl(
		\sum_{i=0}^\infty
		x \cdot \partial
		\partialprod{\sigma}{1}{i}
		\dfield
		+
		\sum_{i=1}^\infty
		\bigl[ \partialprod{\sigma}{1}{i}, x\cdot\partial \bigr] \dfield
	\biggr)
	\pder{\Ltot}{(\partialprod{\sigma}{1}{i} \dfield)}
.
\end{align}
The first term can be processed by the chain rule to give $-\D \Stot$. It is thus apparent that, upon integration, \eq{ClassicalConstraint} reduces to
\be
	\dil{\delta} \dfield \cdot \fder{\Stot}{\dfield} = 0,
\label{eq:ClassicalDilInv}
\ee
which is of course nothing but the statement of dilatation invariance. 

Returning to~\eq{ClassicalConstraint}, we now multiply by $2x_\mu$ and then integrate.
To see what this gives, observe that
\begin{align}
	\sct{0}{\mu} \dfield \cdot \fder{\Stot}{\dfield}
& =
	\Int{x}
	\biggl(
		\sum_{i=0}^\infty \bigl( 2 x_\mu x \cdot \partial - x^2 \partial_\mu \bigr)
		\partialprod{\sigma}{1}{i} \dfield
		+
		\sum_{i=1}^\infty
		\bigl[
			\partialprod{\sigma}{1}{i}
		,
			2 x_\mu x \cdot \partial - x^2 \partial_\mu
		\bigr]
		\dfield
	\biggr)
	\pder{\Ltot}{(\partialprod{\sigma}{1}{i} \dfield)}
\nonumber
\\
& =
	-2d \Int{x} x_\mu \Ltot
	-
	\bigl(
		\delta_{\omega\mu} \delta_{\rho\sigma} 
		- 2\delta_{\omega\rho}\delta_{\sigma\mu}
	\bigr)
	\Int{x}
	\sum_{i=1}^\infty
	\bigl[
		\partialprod{\sigma}{1}{i}
	,
		x_\sigma x_\rho \partial_\omega
	\bigr]
	\dfield
	\pder{\Ltot}{(\partialprod{\sigma}{1}{i} \dfield)}
.
\label{eq:ClassicalSCT-Expand}
\end{align}
The final term can be processed by manipulating the commutator:
\begin{align}
	\bigl[
		\partialprod{\sigma}{1}{i}
	,
		x_\sigma x_\rho \partial_\omega
	\bigr]
&	=
	\bigl[
		\partialprod{\sigma}{1}{i}
	,
		x_\sigma 
	\bigr]
	x_\rho \partial_\omega
	+
	x_\sigma
	\bigl[
		\partialprod{\sigma}{1}{i}
	,
		x_\rho
	\bigr]
	\partial_\omega
\nonumber
\\
& =
	\bigl[
		\bigl[
			\partialprod{\sigma}{1}{i}
		,
			x_\sigma 
		\bigr]
	,
		x_\rho
	\bigr]
	\partial_\omega
	+
	x_\rho
	\bigl[
		\partialprod{\sigma}{1}{i}
	,
		x_\sigma 
	\bigr]
	\partial_\omega
	+
	x_\sigma
	\bigl[
		\partialprod{\sigma}{1}{i}
	,
		x_\rho
	\bigr]
	\partial_\omega
.
\end{align}
From this it follows that
\begin{multline}
	\bigl(
		\delta_{\omega\mu} \delta_{\rho\sigma} 
		- 2\delta_{\omega\rho}\delta_{\sigma\mu}
	\bigr)
	\bigl[
		\partialprod{\sigma}{1}{i}
	,
		x_\sigma x_\rho \partial_\omega
	\bigr]
	=
	2x_\rho 
	\bigl[
		\partialprod{\sigma}{1}{i}, x_\rho \partial_\mu - x_\mu \partial_\rho
	\bigr]
	+
\\
	\bigl(
		\delta_{\omega\mu} \delta_{\rho\sigma} 
		- 2\delta_{\omega\rho}\delta_{\sigma\mu}
	\bigr)
	\bigl[
		\bigl[
			\partialprod{\sigma}{1}{i}
		,
			x_\sigma 
		\bigr]
	,
		x_\rho
	\bigr]
	\partial_\omega
	-2 x_\mu
	\bigl[
		\partialprod{\sigma}{1}{i}
	,
		x \cdot \partial
	\bigr]
.
\end{multline}
When substituted into~\eq{ClassicalSCT-Expand}, the first term on the \rhs\ vanishes as a consequence of~\eq{L-RotSymmetry}; returning to~\eq{ClassicalConstraint}, we thus conclude that multiplying by $x_\mu$ and integrating implies:
\be
	\sct{\delta}{\mu} \dfield \cdot \fder{\Stot}{\dfield} = 0
\label{eq:ClassicalSCTInv}
\ee
which, in combination with~\eq{ClassicalDilInv},
shows that~\eq{ClassicalConstraint} encodes invariance under both dilatations and special conformal transformations.

Supposing that a conformally invariant action has been found, \eq{ClassicalConstraint} does not uniquely determine $\hat{L}$ and $\tilde{h}_{\tau\lambda}$. However, as we shall illustrate in the next section with some concrete examples, the energy-momentum tensor does not depend on the particular choice. It should be possible to demonstrate this invariance generally, though it is beyond the scope of this paper to do so.

\subsubsection{Examples}

\paragraph{Gaussian Theory} To get a feeling for the construction of the energy-momentum tensor including, in particular~\eq{ClassicalConstraint}, consider the case of the Gaussian fixed-point. To start with, take $\Ltot = \hf \partial_\mu \dfield\, \partial_\mu \dfield$. Referring back to~\eqs{t_ab}{f_lab} it is apparent that
\be
	\symmetric{\conserved{\CEMT}_{\alpha\beta}} = 
	\partial_\alpha \dfield \, \partial_\beta \dfield
	- \smallhf \delta_{\alpha \beta} \partial_\mu \dfield\, \partial_\mu \dfield,
	\qquad
	f_{\lambda \alpha \beta} = 0
.
\ee
From~\eq{h'} we see that $h_{\tau\lambda} = \remainder{h}_{\tau\lambda}$, with the latter determined by~\eq{ClassicalConstraint}
\begin{align}
	-\partial_\lambda \partial_\tau \remainder{h}_{\tau\lambda}
	& = 
	-\delta_0 \partial_\lambda \dfield \, \partial_\lambda \dfield 
	- \delta \dfield \,\partial^2 \dfield
\nonumber
\\
	& =
	-\frac{\delta}{2} \partial^2 \dfield^2
	+ \bigl(\delta -\delta_0\bigr) \partial_\lambda \dfield \, \partial_\lambda \dfield,
\end{align}
therefore implying that $\delta = \delta_0$, as expected, with
\be
	h_{\tau\lambda} = \remainder{h}_{\tau\lambda} = \frac{\D-2}{4} \delta_{\tau\lambda} \dfield^2.
\ee
To construct the full energy-momentum tensor we employ~\eqs{Pol-condition-d>2}{Pol-condition-d=2}; conveniently, for the present case where $h_{\tau\lambda} \sim \delta_{\tau\lambda}$, the former reduces to the latter and so we find, using~\eq{symmetricEMT}
\be
	\CEMT_{\alpha \beta}
	=
	\partial_\alpha \dfield \, \partial_\beta \dfield
	- \hf \delta_{\alpha \beta}  \partial_\mu \dfield\, \partial_\mu \dfield
	+\frac{\D-2}{4(\D-1)}
	\bigl(
		\delta_{\alpha\beta}\partial^2 - \partial_\alpha \partial_\beta
	\bigr)
	\dfield^2
,
\ee
as expected.
It is easy to check that taking $\Ltot = -\hf \dfield \, \partial^2 \dfield$ gives the same result and so, reassuringly, $t_{\alpha\beta}$ is independent of which of the representatives of $\Ltot$ we use.

\paragraph{Higher Derivative Theory} As a slightly more involved example, we will explore the free theory with a kinetic term quartic in derivatives. To begin with, we shall consider
$\Ltot = \hf \partial_{\mu} \partial_\nu \dfield\, \partial_{\mu} \partial_\nu \dfield$. Power counting informs us that the scaling dimension of $\dfield$ is $(\D-4)/2 < \delta_0$ and so, in Minkowski space, the theory is non-unitary. Nevertheless, at least for $\D>2$, we can construct the energy-momentum tensor.
Referring back to~\eqs{t_ab}{f_lab}, we have:
\begin{subequations}
\begin{align}
	\symmetric{\conserved{\CEMT}_{\alpha\beta}}  
	& = 
	- \smallhf \delta_{\alpha\beta}  \partial_{\mu} \partial_\nu \dfield \,\partial_{\mu} \partial_\nu \dfield
	+ 2 \partial_{\alpha} \partial_\nu \dfield \, \partial_{\beta} \partial_\nu \dfield
,
\\
	f_{\lambda \alpha \beta}
	&=
	-\partial_\beta \dfield
	\,
	\partial_\alpha \partial_\lambda \dfield
,
\label{eq:partial4-f_lab}
\end{align}
\end{subequations}
corresponding, according to~\eq{TbarSym}, to the conserved but not symmetric tensor
\be
	\conserved{\CEMT}_{\alpha\beta}
	=
	\partial_\mu \partial_\alpha \dfield \, \partial_\mu \partial_\beta \dfield
	-\smallhf \delta_{\alpha\beta} \partial_\mu \partial_\nu \dfield \, \partial_\mu \partial_\nu \dfield
	-\partial_\beta \dfield \, \partial_\alpha \partial^2 \dfield
.
\ee
Following the recipe~\eq{symmetricEMT} yields the conserved, symmetric tensor
\be
	\CEMT_{\alpha\beta}
	=
	\partial_\mu \dfield \, \partial_\mu \partial_\alpha \partial_\beta \dfield
	+ \partial^2 \dfield \, \partial_\alpha \partial_\beta \dfield
	-\smallhf \delta_{\alpha\beta} \partial_\mu \partial_\nu \dfield \, \partial_\mu \partial_\nu \dfield
	-\partial_\beta \dfield \, \partial_\alpha \partial^2 \dfield
	-\partial_\alpha \dfield \, \partial_\beta \partial^2 \dfield
	+\partial_\lambda w_{\lambda \alpha \beta},
\label{eq:partial4-t_ab}
\ee
where $\partial_\lambda w_{\lambda \alpha \beta}$ is determined by~\eq{Pol-condition-d>2} in terms of $h_{\tau\lambda}$. From~\eq{h'}, in this particular case $h_{\tau\lambda} = \remainder{h}_{\tau\lambda}$ and from~\eq{ClassicalConstraint}
\be
	-\partial_\lambda \partial_\tau \remainder{h}_{\tau\lambda}
	=
	\frac{4-\D}{2}
	\partial_{\mu} \partial_\nu \dfield\, \partial_{\mu} \partial_\nu \dfield
	+
	\delta \dfield \, \partial^4 \dfield
	+
	\partial_\mu 
	\bigl(
		\partial_\mu \dfield \, \partial^2 \dfield
		-2 \partial_\nu \dfield \, \partial_\mu \partial_\nu \dfield
	\bigr)
.
\ee
Components of the first two terms can be transferred to the final term by writing
\begin{subequations}
\begin{align}
	\partial_{\mu} \partial_\nu \dfield\, \partial_{\mu} \partial_\nu \dfield
	& =
	\partial_{\mu}
	\bigl(
		\partial_\nu \dfield\, \partial_{\mu} \partial_\nu \dfield
	\bigr)
	-
	\partial_\mu \dfield\, \partial_\mu \partial^2 \dfield
,
\\
	\dfield \, \partial^4 \dfield
	&=
	\partial_\mu
	\bigl(
		\dfield \, \partial_\mu \partial^2 \dfield
	\bigr)
	-
	\partial_\mu \dfield\, \partial_\mu \partial^2 \dfield
\label{eq:massage}
\end{align}
\end{subequations}
so that we have
\be
	-\partial_\lambda \partial_\tau \remainder{h}_{\tau\lambda}
	=
	-\frac{2+\eta}{2} \partial_\mu \dfield \, \partial_\mu \partial^2 \dfield
	+ 
	\partial_\mu
	\bigl(
		\partial_\mu \dfield \, \partial^2 \dfield
		-\D/2 \, \partial_\nu \dfield \, \partial_\mu \partial_\nu \dfield
		+ \delta \dfield \, \partial_\mu \partial^2 \dfield
	\bigr)
.
\ee
The final term can be massaged to give
\[
	\partial_\mu \partial_\nu
	\bigl(
		\partial_\mu \dfield \, \partial_\nu \dfield + \delta \dfield \, \partial_\mu \partial_\nu \dfield
	\bigr)
	-\frac{2\D+ \eta}{4} \partial^2 \bigl(\partial_\mu \dfield \, \partial_\mu \dfield \bigr)
,
\]
and so we conclude that $\eta = -2$, as expected, and, for $\D>2$,
\be
	h_{\tau \lambda} = \remainder{h}_{\tau \lambda}
	=
	\frac{\D-1}{2} \delta_{\tau \lambda} \partial_\mu \dfield \, \partial_\mu \dfield
	- \partial_\lambda \dfield \,\partial_\tau \dfield
	- \frac{\D-4}{2} \dfield \, \partial_\lambda \partial_\tau \dfield
.
\label{eq:partial4-h_tl}
\ee
Unlike in the Gaussian case, the dimension of $\dfield$ is such that we could add an arbitrary transverse term $(\delta_{\tau\lambda}\partial^2 - \partial_\tau \partial_\lambda) \dfield^2$; however, as discussed under~\eq{ConstraintGeneral}, the energy-momentum tensor is insensitive to such contributions.
By inspection of~\eq{Pol-condition-d=2} it is apparent that the solution for $h_{\tau \lambda}$ does not exist in $\D=2$. Before constructing the full energy-momentum tensor, we may perform a simple, intermediate consistency check. Recalling~\eqs{traceOfEMT}{W-extra} and using~\eqss{partial4-f_lab}{partial4-t_ab}{partial4-h_tl} it is straightforward to check that
\be
	\CEMT_{\alpha \alpha}
	=
	-\frac{\D-4}{2} \dfield \, \partial^4 \dfield,
\label{eq:t_aa-Wegner}
\ee
as expected from~\eq{GenDilCondition}. Constructing $\partial_\lambda w_{\lambda \alpha \beta}$ from~\eq{Pol-condition-d>2}, substituting into~\eq{partial4-t_ab} and simplifying yields the full (if rather unwieldy) energy-momentum tensor:
\begin{multline}
	\CEMT_{\alpha \beta}
	=
	\frac{1}{2(\D-2)(\D-1)}
	\Bigl\{
		(4\D-8) \partial_\mu \dfield \, \partial_\mu \partial_\alpha \partial_\beta \dfield
		+\D(\D+2) \partial^2 \dfield \, \partial_\alpha \partial_\beta \dfield
\\
		+(\D-2)(\D-4) \dfield \, \partial_\alpha \partial_\beta \partial^2 \dfield
		+(4-\D^2)
		\bigl(
			\partial_\alpha \dfield \, \partial_\beta \partial^2 \dfield
			+ \partial_\beta \dfield \, \partial_\alpha \partial^2 \dfield
		\bigr)
		-4\D \partial_\alpha \partial_\mu \dfield \, \partial_\beta \partial_\mu \dfield
\\
		+\delta_{\alpha \beta}
		\bigl[
			4 \partial_\mu \partial_\nu \dfield \, \partial_\mu \partial_\nu \dfield
			+(2\D -4)
			\partial_\mu \dfield \, \partial_\mu \partial^2 \dfield
			-(\D+2) \partial^2 \dfield \, \partial^2 \dfield
			-(\D-2)(\D-4) \dfield \, \partial^4 \dfield
		\bigr]
	\Bigr\}
.
\label{eq:HigherDeriv-Full}
\end{multline}
The expression is manifestly symmetric and it is easy to check that it is conserved and traceless (as usual up to terms which vanish on the equations of motion).

A variant of the above analysis which exercises all terms involved in the construction of the energy-momentum tensor is achieved by taking instead $\Ltot = -\hf \partial_\mu \dfield \, \partial_\mu \partial^2 \dfield$. In this case we find:
\begin{subequations}
\begin{align}
	\symmetric{\conserved{\CEMT}_{\alpha\beta}}  
	& = 
	\hf \delta_{\alpha \beta} \partial_\mu \dfield \, \partial_\mu \partial^2 \dfield
	-\hf \partial_\alpha \dfield \, \partial_\beta \partial^2 \dfield
	-\hf \partial_\beta \dfield \, \partial_\alpha \partial^2 \dfield
	-\partial_\mu \dfield\, \partial_\mu \partial_\alpha \partial_\beta \dfield
	,
\\
	f_{\lambda \alpha \beta}
	&=
	-\frac{1}{3} \partial_\beta \dfield \, \partial_\lambda \partial_\alpha \dfield
	-\frac{1}{6} \delta_{\alpha \lambda} \partial_\beta \dfield \, \partial^2 \dfield
	+\frac{1}{3} \partial_\lambda \partial_\beta \dfield \, \partial_\alpha \dfield
	+\frac{1}{3} \partial_\alpha \partial_\beta \dfield \, \partial_\lambda \dfield
	+\frac{1}{3} \delta_{\alpha \lambda} \partial_\mu \partial_\beta \dfield \, \partial_\mu \dfield
;
\end{align}
\end{subequations}
it can be checked that the sum of these terms, ${\conserved{\CEMT}_{\alpha\beta}}$ is conserved.
According to the recipe~\eq{symmetricEMT}, we construct the conserved, symmetric tensor
\begin{multline}
	\CEMT_{\alpha \beta}
	=
	\frac{1}{6} \delta_{\alpha \beta} \partial^2 \dfield \, \partial^2 \dfield
	+\frac{1}{3} \delta_{\alpha \beta} \partial_\mu \dfield \, \partial_\mu \partial^2 \dfield
	 +\frac{2}{3} \partial_\alpha \partial_\beta \dfield \, \partial^2 \dfield
	 +\frac{2}{3} \partial_\mu \dfield \, \partial_\mu \partial_\alpha \partial_\beta \dfield
\\
	 -\frac{1}{3} \delta_{\alpha \beta} \partial_\mu \partial_\nu \dfield \, \partial_\mu \partial_\nu \dfield
	 -\partial_\alpha \dfield \, \partial_\beta \partial^2 \dfield
	 -\partial_\beta \dfield \, \partial_\alpha \partial^2 \dfield
	 +\partial_\lambda w_{\lambda \alpha \beta}
.
\label{eq:HigherDeriv-ConsSymm}
\end{multline}
Contrary to the previous analysis, $h_{\tau\lambda}$ is non-trivially related to $\remainder{h}_{\tau\lambda}$. From~\eq{h'},
\be
	h_{\tau\lambda}
	=
	\remainder{h}_{\tau\lambda}
	- \frac{\D-2}{6} \partial_\tau \dfield \, \partial_\lambda \dfield
	+\frac{1}{3} \delta_{\tau\lambda} \partial_\mu \dfield \, \partial_\mu \dfield
\ee
and from~\eq{ClassicalConstraint}
\be
	-\partial_\lambda \partial_\tau \remainder{h}_{\tau\lambda}
	=
	\frac{\D-4}{2}
	\partial_\mu \dfield \, \partial_\mu \partial^2 \dfield
	+ \delta \dfield \, \partial^4 \dfield
	+
	\partial_\mu
	\bigl(
		\partial_\mu \dfield \, \partial^2 \dfield -
		 \delta_0 \partial_\nu \dfield \, \partial_\mu \partial_\nu \dfield
	\bigr)
.
\ee
Utilizing~\eq{massage},
\be
	-\partial_\lambda \partial_\tau \remainder{h}_{\tau\lambda}
	=
	-\frac{2+\eta}{2}
	\partial_\mu \dfield \, \partial_\mu \partial^2 \dfield
	+
	\partial_\mu
	\bigl(
		\partial_\mu \dfield \, \partial^2 \dfield -
		 \delta_0 \partial_\nu \dfield \, \partial_\mu \partial_\nu \dfield
		 +\delta \dfield \, \partial_\mu \partial^2 \dfield
	\bigr)
\ee
and so, as before, $\eta = -2$. After a bit of rearrangement (and utilizing the result for $\eta$) we find, for $\D>2$:
\be
	\remainder{h}_{\tau \lambda}
	=
	\frac{\D-2}{2} \delta_{\tau \lambda} \partial_\nu \dfield \, \partial_\nu \dfield
	- \partial_\lambda \dfield \,\partial_\tau \dfield
	- \frac{\D-4}{2} \dfield \, \partial_\lambda \partial_\tau \dfield
,
\ee
which yields
\be
	h_{\tau \lambda}
	=
	\frac{3\D-4}{6} \delta_{\tau \lambda} \partial_\mu \dfield \, \partial_\mu \dfield
	-\frac{\D+4}{6} \partial_\tau \dfield \, \partial_\lambda \dfield
	-\frac{\D-4}{2} \dfield \, \partial_\tau \partial_\lambda \dfield
.
\ee
As before, it is relatively easy to confirm~\eq{t_aa-Wegner} but somewhat involved to reconstruct the full energy-momentum tensor~\eq{HigherDeriv-Full}.

\paragraph{Interacting Theories}

A crucial feature of the equations which define the classical energy momentum tensor, \eqss{divT-C}{rotT-C}{dilT-C}, is that contributions to the action depending on powers of the field decouple from one another. However, consistency between these various terms is enforced by the final condition encoding dilatation invariance. For example, consider adding a potential term to the gaussian theory:
\be
	\Ltot = \partial_\mu \dfield \, \partial_\mu \dfield + V(\dfield)
.
\ee
The energy-momentum tensor for this theory is given by:
\be
	\CEMT_{\alpha \beta} = \CEMT_{\alpha\beta}^{\mathrm{Gauss}} + \delta_{\alpha\beta} V(\dfield),
\ee
with dilatation invariance requiring that, for $\D > 2$,
\be
	V(\dfield) \propto \dfield^{\D / \delta_0}
.
\ee

This is very different from the quantum case, where the \rhs\ of the analogues of~\eqs{divT-C}{dilT-C} are quadratic both in the action and also functional derivatives, making the problem of finding explicit solutions very much more difficult.

\subsection{ERG Representation}
\label{sec:ERG}

\subsubsection{Notation and Conventions}

To formulate the ERG equation, we introduce an ultraviolet cutoff function which, as in the introduction, we denote by $K(x,y)$. As with all ingredients of a good ERG equation this function must be quasi-local (\cf\ the discussion below~\eq{Lhat}). Concretely, for coefficients $k_i$, we may write
\be
	K(x,y) = \sum_{i=0}^\infty k_i \; (-\partial^2)^i \DD{\D}{(x-y)}
	=
	\sum_{i=0}^\infty k_i \; (-\partial^2)^{i+1} \ep_0\bigl((x-y)^2\bigr)
.
\label{eq:K(x,y)}
\ee
where $\ep_0$ is Green's function, so that $-\partial^2 \ep_0 = \one$.

From the cutoff function we construct an object $G$ satisfying
\be
	\bigl(
		\D + x\cdot \partial_x + y \cdot \partial_y
	\bigr)
	\cutoff (x,y)
	= \partial^2_x G(x,y)
.
\label{eq:G}
\ee
This is perhaps more intuitive in momentum space%
\footnote{We use the same symbol for a function of coordinates and its Fourier transform.
} where it translates to $p \cdot \partial_p K = p^2 G$ or just $G(p^2) = 2 \,d K(p^2)/dp^2$. From $G$ it is helpful to construct
\be
	G_\mu(x,y) \equiv  (x+y)_\mu G(x,y)
.
\label{eq:G_mu}
\ee

It is useful to define an ultraviolet regulated version of Green's function:
\be
	\ep = \ep_0 \cdot \cutoff
\label{eq:RegProp}
\ee
where, as alluded to in the introduction, we use the following shorthand for integrals:
\be
	\Psi \cdot \Phi
	\equiv
	\Int{x} \Psi(x) \Phi(x),
	\qquad
	\Psi \cdot F \cdot \Phi
	\equiv
	\Int{x}\volume{y} \Psi(x) F(x,y) \Phi(y)
.
\ee
Now we have all the ingredients we need;  the ERG equation and its partner encoding special conformal invariance read, up to vacuum terms (which are neglected throughout this paper):
\begin{subequations}
\begin{align}
	\biggl\{
		\dil{\delta} \dfield \cdot \fder{}{\dfield}
		+ \dfield \cdot \ep^{-1} \cdot G \cdot \fder{}{\dfield}
		+ \hf \fder{}{\dfield} \cdot G \cdot \fder{}{\dfield}
	\biggr\}
	e^{-\Stot}
&
	=
	0,
\label{eq:ERGE}
\\
	\biggl\{
		\sct{\delta}{\mu} \dfield \cdot \fder{}{\dfield}
		+ \dfield \cdot \ep^{-1} \cdot G_\mu \cdot \fder{}{\dfield}
		+ \hf \fder{}{\dfield} \cdot G_\mu \cdot \fder{}{\dfield}
		- \eta \partial_\mu \dfield \cdot \cutoff^{-1} \cdot G \cdot \fder{}{\dfield}
	\biggr\}
	e^{-\Stot}
&
	=
	0.
\label{eq:ERGE-SCT}
\end{align}
\end{subequations}
The ERG equation is slightly different from that usually appearing in the literature---which is the variant of the Wilson/Polchinski equations~\cite{Wilson,Pol} proposed in~\cite{Ball}---on account of it using the full Wilsonian effective action; the relationship to the more common form is given in \app{ERG}. The special conformal equation~\eq{ERGE-SCT} is a similar re-expression of the equation written down in~\cite{Representations}.

The analysis of the following section will utilize some new notation. To motivate this, let us anticipate that in the ERG treatment of the energy-momentum tensor we will encounter a term like
\be
	\partial_\beta \dfield \cdot \cutoff^{-1} \times \cutoff \cdot \fder{\Stot}{\dfield}
	=
	\partial_\beta \dfield \times \fder{\Stot}{\dfield}
	-
	\partial_\beta \dfield \cdot \cutoff^{-1} \cdot
	\bigl(
		\cutoff \times \one - \one \times \cutoff 
	\bigr)
	\cdot \fder{\Stot}{\dfield}
\label{eq:example}
\ee
where given $A(x,y)$ and $B(x,y)$ we understand
\be
	(A \times B)(y,z;x) = A(y,x) B(x,z)
.
\ee
The first term on the \rhs\ of~\eq{example} we recognize from the classical analysis and so our task will be to process the second term. In particular, we would like to re-write it as a total derivative. To this end observe that, for some quasi-local $F\bigl((x-y)^2\bigr)$,
\be
	F \times \one - \one \times F = \partial_\alpha \core{F}_\alpha
,
\label{eq:partial_aF_a}
\ee
where $\core{F}_\alpha(y,z;x)$ is also quasi-local and the partial derivative on the \rhs\ is understood to be \wrt\ $x$. This is easy to see by making the coordinate dependence explicit and integrating:
\[
	\Int{x} \bigl( F(y,x) \DD{\D}(x-z) - \DD{\D}(y-x) F(x, z) \bigr) = 0
\]
and, since $F$ is quasi-local, \eq{partial_aF_a} follows.
Note that, courtesy of translation invariance, $\core{F}$ can be rewritten as a function of coordinate differences: $\core{F}_\alpha\bigl(y-x,z-x\bigr)$. Here we are overloading notation so that the two argument form of $\core{F}$ is considered separate from the three argument form.

Were we to directly utilize~\eq{partial_aF_a} in~\eq{example}, our notation would be potentially confusing since up until now a single object sandwiched between dots---such as $\cdot G \cdot$ in the ERG equation---is such that all coordinates are integrated over. To avoid possible confusion as to whether or not expressions are fully integrated, without having to pollute all of our equations with explicit coordinate dependence, we develop some new notation which retains the compactness of the dot notation. Given test functions $\Psi$ and $\Phi$, we hijack the symbols $\rtimes$ and $\ltimes$ as follows:
\be
	\bigl( \lrtimes{\Psi}{\core{F}}{\Phi} \bigr)(x) \equiv
	\Int{y} \volume{z} \Psi(y) \core{F}(y,z;x) \Phi(z)
.
\label{eq:lrtimes}
\ee
The \lhs\ could be written as $\Psi \cdot \core{F}(x) \cdot \Phi$, where the coordinates of $\core{F}$ which are integrated over have been suppressed. 
However, whereas we should retain the $x$ in this case to avoid ambiguity, we will be able to use the new notation with the explicit coordinate dependence dropped. Amongst other things, this allows such terms such as $\lrtimes{\Psi}{\core{F}}{\Phi}$ to cleanly appear in the same expression as e.g.\ $\Psi \cdot \cutoff \times \Phi$.

Overloading notation so that the same symbol is used for both a function of position and its Fourier transform, and using $k,p$ to denote momenta, we have:
\be
	\bigl( \lrtimes{\Psi}{\core{F}}{\Phi} \bigr)(k) \equiv
	\MomInt{p} \Psi(k-p) \core{F}(k-p,p) \Phi(p)
.
\label{eq:lrtimes-FT}
\ee
The notation of~\eq{lrtimes} naturally extends to the case where $\Phi$ and $\Psi$ depend on two arguments, \viz
\be
	\bigl( \lrtimes{\Psi}{\core{F}}{\Phi} \bigr)(u,v;x) \equiv
	\Int{y} \volume{z} \Psi(u,y) \core{F}(y,z;x) \Phi(z,v)
;
\label{eq:lrtimes-twoArgs}
\ee
clearly it can also be applied to the mixed case where $\Phi$ has one argument but $\Psi$ has two (or vice-versa).

To gain some experience with the new notation, let us record several useful properties:
\begin{subequations}
\begin{align}
	\lrtimes{\Psi}{\partial_\alpha \core{F}}{\Phi}
	&=
	\partial_\alpha \bigl( \lrtimes{\Psi}{\core{F}}{\Phi} \bigr)
\label{eq:lrtimes-totalDeriv}
\\
	\lrtimes{\Psi}{\partial_\alpha \core{F}}{\Phi}
	& =
	\lrtimes{\partial_\alpha \Psi}{\core{F}}{\Phi}
	+\lrtimes{\Psi}{\core{F}}{\partial_\alpha \Phi}
,
\label{eq:PsipartialFPhi}
\\
	\lrtimes{\Psi}{
		\lrtimes{\one}{\core{F}}{\one}
	}{\Phi}
	& =
	\lrtimes{\Psi}{
		\core{F}
	}{\Phi}.
\label{eq:lrtimes-unity}
\end{align}
\end{subequations}
The first equation is a trivial consequence of~\eq{lrtimes}, given the convention stated under~\eq{partial_aF_a} that
$(\partial_\alpha \core{F})(y,z;x) = \partial \core{F}(y,z;x)/\partial x_\alpha$. \Eqn{PsipartialFPhi} exploits translation invariance of $\core{F}$: 
\[
	\Int{y}\volume{z} \Psi(y) \pder{}{x_\alpha} \core{F}(y-x, z-x) \Phi(x)
	=
	-
	\Int{y}\volume{z} \Psi(y) 
	\biggl(
		\pder{}{y_\alpha} 
		+ \pder{}{z_\alpha}	
	\biggr)
	\core{F}(y-x, z-x) \Phi(z)
.
\]
Integrating by parts, the desired result follows.
There are several ways to see~\eq{lrtimes-unity}. Most directly, one could simply substitute $\delta$-functions for the $\one$s and apply~\eq{lrtimes-twoArgs}:
\be
	\lrtimes{\one}{\core{F}}{\one}
	=
	\Int{y}\volume{z} \DD{\D}(u-y) \core{F}(y,z;x) \DD{\D}(z-v)
	=
	\core{F}(u,v;x)
;
\ee
after applying~\eq{lrtimes} again, \eq{lrtimes-unity} follows. At a more heuristic level, one could swap the $\ltimes$ and $\rtimes$ for dots---mindful of the ambiguity above---and then convert them back again after eliding the $\one$s.

The notation of~\eq{lrtimes} allows us to neatly express $\core{F}_\alpha$ in terms of derivatives of a scalar:
\be
	\core{F}_\alpha
	=
	\partial_\alpha  \lrtimes{\one}{\core{F}}{\one} 
	- \lrtimes{\one}{\core{F}}{\one} \lpartial_\alpha 
.
\label{eq:Falpha}
\ee
The backward-pointing arrow indicates that the associated operator---in this case a partial derivative---acts on the last argument of the object to its left, which here would correspond to $\partial \DD{\D}(y,u)/\partial u_\alpha$, with $\DD{\D}(y,u) = \DD{\D}(y-u)$.
To justify~\eq{Falpha},  we establish a relationship between $\core{F}$ and $F$.
Employing~\eq{PsipartialFPhi}, observe that:
\be
	\lrtimes{\Psi}{
		\partial_\alpha \core{F}_\alpha
	}{\Phi}
	=	
	\lrtimes{\Psi}{\core{F}}{\partial^2 \Phi}
	-\lrtimes{\partial^2 \Psi}{\core{F}}{\Phi}
.
\ee
This can be compared with~\eq{partial_aF_a}; it is particularly transparent to do so in momentum space, which yields:
\be
	\core{F}(k-p, p) 
	=
	\frac{F\bigl((k-p)^2\bigr) - F(p^2)}{k\cdot (k-2p)}
.
\ee
The expansion in $k$ will play a key role later on; in its most useful form it is:
\be
	\core{F}(k-p, p)
	=
	\hf
	\bigl[
		F'(p^2) 
		+ F'\bigl((k-p)^2\bigr)
	\bigr]
	+ \order{k^2}
,
\ee
where the prime denotes a derivative \wrt\ the argument.
Equivalently,
\begin{align}
	\lrtimes{\Psi}{\core{F}}{\Phi}
&	= 
	\hf
	\Psi \cdot
	\bigl(
		F' \times \one + \one \times F'
	\bigr)
	\cdot \Phi
	+\order{\partial^2}
\nonumber
\\
	&= 
	\hf
	\Psi \cdot
	\acom{F'}{\one}
	\cdot \Phi
	+\order{\partial^2}
,
\label{eq:F-expand}
\end{align}
where the second line defines the notation $\{ \cdot, \cdot \}$ and, in position space, we understand $F'$ to be the Fourier transform of $d F(p^2) / dp^2$, and so forth. Equivalently, recalling~\eq{dil^delta}, we can define $F'$ via:
\be
	\dil{\D/2} F + F \dilL{\D/2} = 2\partial^2 F'
	\qquad
	\mathrm{or}
	\qquad
	\dil{\delta_0} \ep_0 \cdot F + \ep_0 \cdot F \dilL{\delta_0}  = -2 F'
,
\label{eq:F'}
\ee
where
\be
	\bigl(\dil{\Delta} F\bigr)(x,y) = \bigl(x \cdot \partial_x + \Delta\bigr)F(x,y)
	\qquad
	\bigl(F\dilL{\Delta}\bigr)(x,y) = \bigl(y \cdot \partial_y + \Delta\bigr)F(x,y)
.
\ee
Note that combining~\eqs{G}{F'} leads to the familiar identification
\be
	\cutoff' \equiv G/2
.
\label{eq:K'}
\ee

We conclude this section by giving some useful equations which follow from~\eq{Falpha}.
Integrating over test functions and employing~\eq{lrtimes-unity} yields
\begin{subequations}
\begin{align}
	\lrtimes{\Psi}{\core{F}_\alpha}{\Phi}
	&=
	-\lrtimes{\partial_\alpha \Psi}{\core{F}}{\Phi}
	+\lrtimes{\Psi}{\core{F}}{\partial_\alpha \Phi}
\label{eq:Falpha-testfns}
\\
	&=
	-\partial_\alpha 
	\bigl( \lrtimes{\Psi}{\core{F}}{\Phi} \bigr)
	+ 2 \lrtimes{\Psi}{\core{F}}{\partial_\alpha \Phi}
\label{eq:Falpha-testfns-}
\\
	&=
	\partial_\alpha 
	\bigl( \lrtimes{\Psi}{\core{F}}{\Phi} \bigr)
	- 2 \lrtimes{\partial_\alpha \Psi}{\core{F}}{\Phi}
\label{eq:Falpha-testfns+}
.
\end{align}
\end{subequations}

\subsubsection{Analysis}

To specialize the general analysis of the energy-momentum tensor of \sect{General} to the ERG requires expressions for the $\quasi^{(\delta)}$ and $\hat{\quasi}^{(\D-\delta)}$ appearing in~\eqs{divT}{traceT}. First of all, we note the existence of a pair of primary fields~\cite{Osborn+Twigg,HO-Remarks,Fundamentals,Representations}
\begin{subequations}
\begin{align}
	\quasi^{(\delta)}_{\mathrm{ERG}}
	&
	=
	\cutoff^{-1} \cdot \dfield - \R \cdot \fder{\Stot}{\dfield}
,
\\
	\quasi^{(\D-\delta)}_{\mathrm{ERG}}
	&
	=
	\fder{\Stot}{\dfield} \cdot \cutoff
,
\end{align}
\end{subequations}
where the subscript ERG is a reminder that we are in the (quasi-local) ERG representation and, in momentum space, for $\eta <2$
\be
	\R(p^2) = p^{2(\eta/2-1)} \cutoff(p^2)
	\int_0^{p^2} d q^2
	q^{-2(\eta/2)}
	\der{}{q^2} 
	\frac{1}{\cutoff(q^2)}
.
\label{eq:rho}
\ee
Given that the cutoff function is normalized such that $K(p^2) = 1 + \order{p^2}$ it follows that,
as anticipated earlier, $\quasi^{(\delta)}_{\mathrm{ERG}}$ is, up to cutoff-dependent terms, just $\dfield$, whereas $\quasi^{(\D-\delta)}_{\mathrm{ERG}}$ is similarly related to $\delta S / \delta \dfield$. These equations may be verified by checking that~\eqs{Odelta}{Od-delta} hold with $\Dil$ and $\SCT_\mu$ given, respectively, by~\eqs{Dil-ERG}{Sct-ERG}, so long as we multiply $\quasi^{(\delta)}_{\mathrm{ERG}}$ and $\quasi^{(\D-\delta)}_{\mathrm{ERG}}$ by $e^{-\Stot}$. Equivalently, we can stick with $\quasi^{(\delta)}_{\mathrm{ERG}}$ and $\hat{\quasi}^{(\D-\delta)}_{\mathrm{ERG}}$ and take a representation of the dilatation operator give by $\Dil_{\Stot} = e^{\Stot}\, \Dil\, e^{-\Stot}$~\cite{Representations}.

The extension of $\quasi^{(\D-\delta)}_{\mathrm{ERG}}$ to an $\hat{\quasi}^{(\D-\delta)}_{\mathrm{ERG}}$ satisfying~\eq{hatO} is given simply by~\cite{Representations}
\be
	\hat{\quasi}^{(\D-\delta)}_{\mathrm{ERG}}
	=
	\fder{\Stot}{\dfield} \cdot \cutoff
	-\fder{}{\dfield} \cdot \cutoff
,
\ee
as may be checked using~\eq{hatO}, with the same qualifications as above.

Recall that, in the classical case, by a conserved and traceless energy-momentum tensor we mean that the \rhss\ of~\eqs{divT}{traceT} vanish on the equations of motion. The equivalent statement in the ERG treatment is that the \rhss\ of the equations are `redundant'---with a redundant field defined such that it is generated by quasi-local field redefinition~\cite{WegnerInv,Wegner-CS}. (The term `inessential' is also used in the literature.)

We are now ready to attempt to solve~\eqss{divT}{symT}{traceT} in the ERG representation. The first step is to split the energy-momentum tensor up into a `classical' piece and `quantum' piece:
\be
	T_{\alpha \beta} = \CEMT_{\alpha\beta} + \QEMT_{\alpha \beta}.
\label{eq:split}
\ee

Conservation of the energy-momentum tensor implies:
\be
	\partial_\alpha T_{\alpha \beta}
	=
	e^{\Stot}
	\fder{}{\dfield} \cdot \cutoff
	\times
	\partial_\beta
	\biggl(
		\R \cdot \fder{}{\dfield} + \cutoff^{-1} \cdot \dfield
	\biggr)
	e^{-\Stot}
.
\label{eq:divT-ERG}
\ee
Neglecting a (divergent) vacuum contribution to the classical term, \eq{divT-ERG} decomposes into (\cf~\eq{example})
\begin{subequations}
\begin{align}
	\partial_\alpha \CEMT_{\alpha \beta}
	&
	=
	-\partial_\beta \dfield \times \fder{\Stot}{\dfield}
,
\label{eq:divT-ERG-C}
\\
	\partial_\alpha \QEMT_{\alpha\beta} \, e^{-\Stot}
	& =
	\biggl\{
		\fder{}{\dfield} \cdot \cutoff
		\times \partial_\beta \R \cdot \fder{}{\dfield}
		-
		\partial_\beta \dfield \cdot \cutoff^{-1} \cdot
		\bigl(
			 \cutoff \times \one -  \one \times \cutoff
		\bigr)
		\cdot
		\fder{}{\dfield}
	\biggr\}
	e^{-\Stot}
\label{eq:divT-ERG-Q}
\end{align}
\end{subequations}
where equality in the final equation is strictly only up to a vacuum contribution.
Since the classical part has been treated in \sect{Classical}, we focus on the quantum contribution, the first term of which can be readily re-expressed by exploiting~\eqs{partial_aF_a}{lrtimes-totalDeriv}:
\be
	\partial_\beta \dfield \cdot \cutoff^{-1} \cdot
	\bigl(
		 \cutoff \times \one -  \one \times \cutoff
	\bigr)
	\cdot
	\fder{}{\dfield}
	=
	\partial_\alpha
	\biggl(
		\partial_\beta \dfield \cdot
		\lrtimes{\cutoff^{-1}}{\core{\cutoff}_\alpha}{\fder{}{\dfield}}
	\biggr)
.
\ee
 To process the double derivative term, we again exploit~\eq{partial_aF_a} to pull out a total derivative piece:
\begin{multline}
	\fder{}{\dfield} \cdot \cutoff \times \partial_\beta \R \cdot \fder{}{\dfield}
	=
	\hf
	\partial_\alpha
	\biggl(
		\delta_{\alpha \beta} \fder{}{\dfield} \cdot\cutoff \times \R \cdot \fder{}{\dfield}
\\
		-\fder{}{\dfield} \cdot \lrtimes{\cutoff}{\core{\R}_\alpha}{\partial_\beta \fder{}{\dfield}}
		-\lrtimes{\partial_\beta \fder{}{\dfield}}{\core{\cutoff}_\alpha}{\R} \cdot \fder{}{\dfield}
	\biggr)
,
\label{eq:DoubleFDerTerm}
\end{multline}
where $\core{R}_\alpha$ is to $\R$ what $\core{K}_\alpha$ is to $\cutoff$ is what $\core{F}_\alpha$ is to $F$. Equality follows straightforwardly from expanding out the \rhs\ using~\eqs{lrtimes-totalDeriv}{partial_aF_a} and noting that $\cutoff \cdot \R = \R \cdot \cutoff$ (which is particularly obvious in momentum space).
Thus we can construct the following contribution to the energy-momentum tensor:
\begin{multline}
	\conserved{\QEMT}_{\alpha\beta} e^{-\Stot}
	=
	-\biggl(
		\partial_\beta \dfield \cdot \lrtimes{\cutoff^{-1}}{\core{K}_\alpha}{\fder{}{\dfield}}
		- \hf \delta_{\alpha \beta} \fder{}{\dfield} \cdot\cutoff \times \R \cdot \fder{}{\dfield}
\\
		+\hf\fder{}{\dfield} \cdot \lrtimes{\cutoff}{\core{\R}_\alpha}{\partial_\beta \fder{}{\dfield}}
		+\hf\lrtimes{\partial_\beta \fder{}{\dfield}}{\core{\cutoff}_\alpha}{\R} \cdot \fder{}{\dfield}
	\biggr)
	e^{-\Stot}
.
\label{eq:Qbar_ab}
\end{multline}

To form the symmetric energy-momentum tensor, according to the recipe which ultimately produces~\eq{symmetricEMT}, we use~\eqs{Falpha-testfns-}{Falpha-testfns+} to re-express in terms of symmetric pieces
plus total derivative terms:
\begin{subequations}
\begin{align}
	\lrtimes{\Psi}{\core{F}_\alpha}{\partial_\beta \Phi}
&	=
	-\partial_\alpha
	\Bigl(
		\lrtimes{\Psi}{\core{F}}{\partial_\beta \Phi}
	\Bigr)
	+
	2\lrtimes{\Psi}{\core{F}}{\partial_\alpha \partial_\beta  \Phi}
,
\\
	\lrtimes{\partial_\beta \Psi}{\core{F}_\alpha}{\Phi}
	& =
	\partial_\alpha
	\Bigl(
		\lrtimes{\partial_\beta \Psi}{\core{F}}{\Phi}
	\Bigr)
	-
	2\lrtimes{\partial_\alpha \partial_\beta\Psi}{\core{F}}{\Phi}
.
\end{align}
\end{subequations}
Utilizing the manifest symmetry under $\alpha \leftrightarrow \beta$ of the final terms allows
us to decompose $\conserved{\QEMT}_{\alpha\beta}$ along the lines of~\eq{TbarSym}, with:
\begin{multline}
	\symmetric{\conserved{\QEMT}}_{\alpha \beta} e^{-\Stot}
	=
	\biggl(
		2\partial_\alpha \partial_\beta \dfield \cdot
		\lrtimes{\cutoff^{-1}}{
			\core{\cutoff}
		}{\fder{}{\dfield}}
		+
		 \hf \delta_{\alpha \beta} \fder{}{\dfield} \cdot	
		 \cutoff \times \R \cdot \fder{}{\dfield}
\\
		-\fder{}{\dfield} \cdot 
		\lrtimes{\cutoff}{
			\core{\R}
		}{\partial_\alpha \partial_\beta \fder{}{\dfield}}
		+
		\lrtimes{\partial_\alpha \partial_\beta \fder{}{\dfield}}{
			\core{\cutoff}
		}{R} \cdot \fder{}{\dfield}
	\biggr)
	e^{-\Stot}
\label{eq:QbarSym}
\end{multline}
and
\be
	F_{\lambda \alpha \beta} e^{-\Stot}
	= 
	-\delta_{\lambda \alpha}
	\biggl(
		 \partial_\beta \dfield \cdot \lrtimes{\cutoff^{-1}}{\core{K}}{\fder{}{\dfield}}
		- \hf \fder{}{\dfield} \cdot \lrtimes{\cutoff}{\core{\R}}{\partial_\beta \fder{}{\dfield}}
		+ \hf \lrtimes{\partial_\beta \fder{}{\dfield}}{\core{\cutoff}}{\R} \cdot \fder{}{\dfield}
	\biggr)
	e^{-\Stot} 
.
\label{eq:F_lab}
\ee

As before, the next step is to construct the trace using~\eq{traceOfEMT} and, after adding the classical piece, to compare with the ERG version of~\eq{traceT}. The first step gives:
\begin{multline}
	\QEMT_{\alpha\alpha} = \partial_\lambda \partial_\tau I_{\tau \lambda}
	+
	e^{\Stot}
	\biggl\{
		-\partial_\lambda \dfield \cdot 
		\lrtimes{\cutoff^{-1}}{
			\bigl[\core{K}_\lambda+(1-\D)\partial_\lambda \core{K}\bigr]
		}{\fder{}{\dfield}}
		+\frac{\D}{2}
		\fder{}{\dfield} 
		\cdot \cutoff \times \R
		\cdot
		\fder{}{\dfield}
\\
		-\hf \fder{}{\dfield} \cdot
		\lrtimes{\cutoff}{
			\bigl[
				\core{R}_\lambda
				+(\D-1) \partial_\lambda \core{R}
			\bigr]
		}{\partial_\lambda \fder{}{\dfield}}
		-\hf 
		\lrtimes{\partial_\lambda \fder{}{\dfield}}{
			\bigl[
				\core{\cutoff}_\lambda
				+(1-\D) \partial_\lambda \core{\cutoff}
			\bigr]
		}{\R}
		\cdot \fder{}{\dfield}
	\biggr\}
	e^{-\Stot}
\label{eq:Q_aa}
\end{multline}
where we have utilized~\eq{lrtimes-totalDeriv} and, recalling~\eq{W-extra}, have split $H_{\tau\lambda}$ into classical and quantum pieces:
\be
	H_{\tau\lambda} = h_{\tau \lambda} + I_{\tau \lambda}
.
\label{eq:H=h+I}
\ee

Again, the strategy is to simplify by absorbing $\order{\partial^2}$ pieces into the first term on the \rhs\ of~\eq{Q_aa}.
To proceed we exploit~\eq{Falpha-testfns+} to re-express
\be
	\lrtimes{\Psi}{
		\bigl(\core{F}_\lambda + (1-\D) \partial_\lambda \core{F} \bigr)
	}{\Phi}
	=
	-(\D-2)
	\lrtimes{\Psi}{
		\partial_\lambda \core{F}
	}{\Phi}
	-2
	\lrtimes{\partial_\lambda \Psi}{\core{F}}{\Phi} 
,
\ee
and then utilize~\eqs{lrtimes-totalDeriv}{F-expand}:
\be
	2\delta_0 \lrtimes{\Psi}{\partial_\lambda \core{F}}{\Phi}
	+
	2\lrtimes{\partial_\lambda \Psi}{\core{F}}{\Phi}
	=
	\delta_0
	\partial_\lambda
	\bigl(
		\Psi\cdot
		\acom{F'}{\one}
		\cdot \Phi
	\bigr)
	+
	\partial_\lambda \Psi
	\cdot
	\acom{F'}{\one}
	\cdot \Phi
	+\order{\partial^2}
,
\ee
where we recall that $\delta_0 \equiv (\D-2)/2$.
Exploiting the notation of~\eq{K'} gives, for some $I^{(1)}_{\tau \lambda}$:
\begin{multline}
	-\partial_\lambda \dfield \cdot \cutoff^{-1} \rtimes
	\bigl(
		\core{K}_\lambda
		+(1-\D) \partial_\lambda\core{K}
	\bigr)
	\ltimes\fder{}{\dfield}
	=
	\partial_\lambda \partial_\tau I^{(1)}_{\tau \lambda}
\\
	+
	\frac{\delta_0}{2}
	\partial_\lambda
	\biggl(
		\partial_\lambda \dfield \cdot \cutoff^{-1} 
		\cdot
		\acom{G}{\one}
		\cdot \fder{}{\dfield}	
	\biggr)
	-
	\hf
	\dfield \cdot \ep^{-1} \cdot
	\acom{G}{\one}
	 \cdot \fder{}{\dfield}
\label{eq:singleder}
\end{multline}
where, in the final term on the \rhs, we have used~\eq{RegProp} to set  $\partial^2 \cutoff^{-1} = - \ep^{-1}$.

To treat the double functional derivative terms in~\eq{Q_aa} note that, for quasi-local $A_1$ and $A_2$
\be
	\fder{}{\dfield} \cdot A_1 \times A_2 \cdot \partial_\lambda \fder{}{\dfield} = \order{\partial}
.
\label{eq:Order-d}
\ee
This can be seen by integrating: the integrand of the \lhs\ is odd and so vanishes.%
\footnote{Strictly speaking, this conclusion holds only if the functional derivatives strikes a translationally invariant functional; more generally, a vacuum term may survive, as discussed in~\cite{Representations}.}
This result implies that various quantities of interest are $\order{\partial^2}$.
\begin{subequations}
\begin{align}
	\partial_\lambda
	\biggl(
		\fder{}{\dfield} \cdot A_1 \times A_2 \cdot \partial_\lambda \fder{}{\dfield}
	\biggr)  
	& = \order{\partial^2}
,
\label{eq:Order_d^2-a}
\\
	\partial^2\fder{}{\dfield} \times A_1 \cdot \fder{}{\dfield}
	-
	\fder{}{\dfield} \times A_1 \cdot \partial^2 \fder{}{\dfield}
	 & = \order{\partial^2}
,
\\
	\fder{}{\dfield} \cdot
	 \lrtimes{A_1}{\partial_\alpha \core{F}_\alpha}{A_2}
	 \cdot \fder{}{\dfield}
	 & =
	 \order{\partial^2}
.
\label{eq:lrtimes-O-d^2}
\end{align}
\end{subequations}
The first equation is a trivial consequence of~\eq{Order-d}. The second equation follows from the first: re-writing the first term as a total derivate plus correction, it is apparent that at $\order{\partial}$ only the correction survives. Thus, at this order, we may transfer derivatives from one side of the $\times$ to the other (and since $A_{1,2}$ are translationally invariant, they are effectively transparent to $\partial_\lambda$). Finally, consider~\eq{lrtimes-O-d^2}. From~\eq{lrtimes-totalDeriv} it is manifestly at least $\order{\partial}$. That it is in fact $\order{\partial^2}$ follows from~\eq{Falpha}, integrating by parts, and then applying~\eqss{lrtimes-unity}{F-expand}{Order_d^2-a}.
These considerations, together with $\{G, \one\} = 2 \one \times G + \order{\partial}$, allow us to finesse a number of terms at $\order{\partial^2}$:
\be
	\hf
	\partial_\lambda \fder{}{\dfield}
	\rtimes
	\bigl(		
		\core{K}_\lambda
		+(1-\D) \partial_\lambda \core{K}
	\bigr)
	\ltimes \R
	\cdot
	\fder{}{\dfield}
	=
	-\hf
	\fder{}{\dfield} 
	\times
	G \cdot \R	
	\cdot
	\partial^2
	\fder{}{\dfield}
	-\partial_\lambda \partial_\tau I^{(2)}_{\tau \lambda}.
\label{eq:doubleder-1}
\ee

To treat the final double derivative term in~\eq{Q_aa} we utilize~\eq{Falpha-testfns-} rather than~\eq{Falpha-testfns+} to re-express:
\be
	\lrtimes{\Psi}{
		\bigl(\core{F}_\lambda + (\D-1) \partial_\lambda \core{F} \bigr)
	}{\Phi}
	=
	(\D-2)
	\lrtimes{\Psi}{
		\partial_\lambda \core{F}
	}{\Phi}
	+2
	\lrtimes{\Psi}{\core{F}}{\partial_\lambda\Phi} 
.
\ee
Noting from~\eq{rho} that
\be
	\der{\R}{p^2}
	=
	\frac{1}{p^2}
	\biggl(
		(\eta/2 -1) \R - \frac{G(p^2)}{2\cutoff(p^2)} 
	\biggr)
	+ \frac{G(p^2)}{2\cutoff(p^2)} \R
,
\label{eq:drho/dp^2}
\ee
the double derivative term under consideration can now be processed similarly to the previous case, yielding:
\begin{multline}
	\hf
	\fder{}{\dfield}
	\cdot
	\cutoff
	\rtimes
	\bigl(		
		\core{R}_\lambda 
		+(\D-1) \partial_\lambda \core{R}
	\bigr)
	\ltimes
	\partial_\lambda
	\fder{}{\dfield}
	e^{-\Stot}
\\
	=
	\biggl[
		(1-\eta/2)
		\fder{}{\dfield} \cdot \cutoff
		\times \R \cdot \fder{}{\dfield}
		+\hf \fder{}{\dfield} \cdot G \times \fder{}{\dfield}
		+\hf \fder{}{\dfield} \times G \cdot \R \cdot \partial^2 \fder{}{\dfield} 
		-\partial_\lambda \partial_\tau I^{(3)}_{\tau \lambda} 
	\biggr] e^{-\Stot}
.
\label{eq:doubleder-2}
\end{multline}
Substituting~\eqss{singleder}{doubleder-1}{doubleder-2} into~\eq{Q_aa} gives:
\begin{multline}
	\QEMT_{\alpha \alpha} e^{-\Stot}
	=
	-
	\biggl[
		 \hf\dfield \cdot \ep^{-1} \cdot 
		 \acom{G}{\one}
		 \cdot \fder{}{\dfield}
		+\hf \fder{}{\dfield} \cdot G \times \fder{}{\dfield}
		-\delta \fder{}{\dfield} \cdot \cutoff \times \R \cdot \fder{}{\dfield}
\\
	-
	\frac{\delta_0}{2} \partial_\lambda
	\biggl(
		\partial_\lambda \dfield \cdot \cutoff^{-1} \cdot 
		 \acom{G}{\one}
		\cdot \fder{}{\dfield}
	\biggr)
	-
	\partial_\lambda \partial_\tau 
	\bigl(
		I + I^{(1)} + I^{(2)} + I^{(3)} 
	\bigr)_{\tau \lambda}
	\biggr]
	e^{-\Stot}
.
\label{eq:Qaa}
\end{multline}

For the full quantum field theoretic case, the trace of the energy-momentum tensor is given by:
\be
	T_{\alpha \alpha}
	=
	\delta
	e^{\Stot}
	\fder{}{\dfield} \cdot \cutoff
	\times
	\biggl(
		\R \cdot \fder{}{\dfield} + \cutoff^{-1} \cdot \dfield
	\biggr)
	e^{-\Stot}
.
\label{eq:T_aa-ERG}
\ee
To extract the `quantum' part, we mirror the decomposition of~\eq{divT-ERG} into~\eqs{divT-ERG-C}{divT-ERG-Q}:
\be
	\dfield \cdot \cutoff^{-1} \times \cutoff \cdot \fder{}{\dfield}
	=
	\dfield \times \fder{}{\dfield}
	-
	\partial_\lambda
	\biggl(
		\dfield \cdot
		\lrtimes{\cutoff^{-1}}{
			\core{\cutoff}_\lambda
		}{\fder{}{\dfield}}
	\biggr)
.
\label{eq:extract}
\ee
Putting the classical $\dfield \times \delta/\delta \dfield$ term to one
side, we compare the remainder of~\eq{T_aa-ERG} to~\eq{Qaa}. To facilitate
this, define
\be
	\partial_\lambda \partial_\tau  I^{(4)}_{\tau\lambda} 
	=
	\delta
	e^{\Stot}
	\partial_\lambda
	\biggl[
		\hf
		\partial_\lambda \dfield \cdot \cutoff^{-1} \cdot
		\acom{G}{\one}
		 \cdot \fder{}{\dfield}
		+
		\dfield \cdot \lrtimes{\cutoff^{-1}}{
			\core{\cutoff}_\lambda
		}{\fder{}{\dfield}}
	\biggr]
	e^{-\Stot}
.
\label{eq:I4}
\ee
That the \rhs\ is $\order{\partial^2}$ follows by using~\eq{Falpha-testfns+}, followed by~\eq{F-expand}.
The comparison of quantum terms now yields
\begin{multline}
	\partial_\lambda \partial_\tau  \remainder{I}_{\tau \lambda}
	 e^{-\Stot} + \ldots
	=
	\biggl[
		\hf\dfield \cdot \ep^{-1} \cdot
		\acom{G}{\one}
		\cdot \fder{}{\dfield}
\\
		+ \hf \fder{}{\dfield} \cdot G \times \fder{}{\dfield}
		+
		\frac{\eta}{4}
		\partial_\lambda
		\biggl(
			\partial_\lambda \dfield 
			\cdot 
			\cutoff^{-1}
			\cdot
			\acom{G}{\one}
			\cdot \fder{}{\dfield}
		\biggr)
	\biggl]
	e^{-\Stot}
	+\ldots
,	
\label{eq:QuantumConstraint}
\end{multline}
where the ellipsis on each side represents the omitted classical terms and
\be
	\partial_\lambda \partial_\tau  \remainder{I}_{\tau \lambda}
	=
	\partial_\lambda \partial_\tau 
	\bigl(
		I + I^{(1)} + I^{(2)} + I^{(3)} + I^{(4)}  
	\bigr)_{\tau \lambda}
.
\label{eq:I'}
\ee

Finally, we add into~\eq{QuantumConstraint} the classical contributions from~\eq{ClassicalConstraint} to give the full result:
\begin{multline}
	\partial_\lambda \partial_\tau \remainder{H}_{\tau\lambda}
 = 
	-
	e^{\Stot}
	\biggl\{
		\D\Ltot
		-
		\sum_{i=1}^\infty
		\bigl[ \partialprod{\sigma}{1}{i}, x\cdot\partial \bigr] \dfield
		\times
		\pder{\Ltot}{(\partialprod{\sigma}{1}{i}\dfield)}
\\
		+
		\delta \dfield \times \fder{}{\dfield}
		+ 
		\hf
		\dfield \cdot \ep^{-1} 
		\cdot 
		\acom{G}{\one}
		\cdot
		\fder{}{\dfield}
		+ 
		\hf \fder{}{\dfield} \cdot G \times \fder{}{\dfield}
\\
	+\partial_\lambda
	\biggl(
		\hf
		\bigl(
			\delta_{\omega \lambda} \delta_{\rho\sigma}
			- 2\delta_{\omega \rho} \delta_{\sigma\lambda}
		\bigr)
		\sum_{i=2}^\infty
		\bigl[
			\bigl[
				\partialprod{\sigma}{1}{i}, x_\sigma
			\bigr],
			x_\rho \partial_\omega
		\bigr]
		\dfield
		\times
		\pder{\Ltot}{(\partialprod{\sigma}{1}{i} \dfield)}
\\
		+
		\frac{\eta}{4}
		\partial_\lambda \dfield 
		\cdot 
		 \cutoff^{-1} \cdot
		 \acom{G}{\one}
		\cdot \fder{}{\dfield}
	\biggr)
	\biggr\}
	e^{-\Stot}
\label{eq:CERG}
\end{multline}
where~\eq{condensed_partials} is employed. 

We can check consistency of~\eq{CERG} just as we did in the classical case: first, we integrate the equation as it stands. The total derivative terms vanish; the first three terms on the \rhs\ have already been dealt with in the reduction of~\eq{ClassicalConstraint} to~\eq{ClassicalDilInv} and it is easy to see that they combine with the remaining terms combine to give  the ERG equation~\eq{ERGE}.

Returning to ~\eq{CERG}, now we multiply by $2 x_\mu$ and then integrate. In this case, all four `classical' terms have been processed in the reduction of~\eq{ClassicalConstraint} to~\eq{ClassicalSCTInv}. Now the various terms combine, straightforwardly, to give the special conformal partner of ERG equation, \eq{ERGE-SCT}.

Finally, given a solution to~\eq{CERG}, the energy-momentum tensor may be reconstructed as follows. First, recall that a recipe for the energy-momentum tensor is provided by~\eq{symmetricEMT}, together with~\eqss{TbarSym}{Pol-condition-d>2}{Pol-condition-d=2}. It was convenient to decompose $\symmetric{\overline{T}}_{\alpha\beta}$ and $F_{\lambda\alpha\beta}$ into classical contributions, given by~\eqs{t_ab}{f_lab}, and quantum ones, \eqs{QbarSym}{F_lab}. The improvement term follows from the decomposition~\eq{H=h+I}, together with~\eqs{h'}{I'}:
\be
	H_{\tau\lambda} 
	= \remainder{H}_{\tau\lambda} + \bigl(I-\remainder{I}\bigr)_{\tau \lambda}
 	+ \bigl(h-\remainder{h}\bigr)_{\tau \lambda}
.
\label{eq:H}
\ee
The various numbered contributions on the \rhs\ of~\eq{I'} may be extracted from~\eq{singleder}, \eq{doubleder-1}, \eqs{doubleder-2}{I4}.

\section{Conclusion}
\label{sec:Conc}

The starting point for the main analysis of this paper is the defining equations for the energy-momentum tensor~\eqss{divT}{symT}{traceT}, written in an arbitrary representation of the conformal algebra.  These three equations respectively encode translation, rotation and dilatation invariance. Supposing that the first constraint can be solved, it is possible to solve the second also to arrive at a conserved, symmetric tensor along the lines of the Belinfante tensor. However, additionally imposing dilatation invariance produces a constraint equation~\eq{ConstraintGeneral}; solutions of this equation, should they exist, provide the requisite improvement to the energy-momentum tensor, while self consistently determining the action and scaling dimension of the fundamental field.

This scheme for constructing the energy-momentum tensor is explored in two concrete representations of the conformal algebra. The treatment of classical theories has a rather standard feel, with the novelty---such as there is one---arising from allowing the Lagrangian to contain terms with an arbitrary number of derivatives. The motivation for this is not that we are interested in classical theories of this type, per se, but rather that such terms necessarily arise in the full ERG treatment which follows. The classical analysis also provides a relatively simple setting in which to directly see that the constraint equation~\eq{ConstraintGeneral} directly encodes both dilatation and special conformal invariance of the action. 

For a classical CFT, the energy-momentum tensor can be reconstructed as follows: a conserved, symmetric tensor can be built from~\eqs{tbar_ab}{f_lab} using the recipe~\eq{symmetricEMT} but excluding the final term. The latter improvement is determined by substituting the solution to the constraint equation~\eq{ClassicalConstraint} into~\eq{h'} and then substituting the result into either~\eq{Pol-condition-d>2} or~\eq{Pol-condition-d=2}, as appropriate. 

The ERG analysis is facilitated by the new notation introduced in~\eq{lrtimes}. This makes it relatively straightforward to solve the conservation equation~\eq{divT} which, in the ERG representation, translates to~\eq{divT-ERG}. The solution separates into classical and quantum pieces, with the former dealt with already and the latter given by~\eq{Qbar_ab}. The new notation swiftly enables the extraction of $F_{\lambda \alpha \beta}$ in~\eq{F_lab}; at this stage a conserved, symmetric tensor could be readily constructed, again by using~\eq{symmetricEMT} modulo the last term. The final part of the analysis involves the improvement of the energy-momentum tensor. The goal of obtaining the ERG representation of the constraint equation~\eq{ConstraintGeneral} is achieved in the `conformal fixed-point equation', \eq{CERG}---the exploration of which is deferred to future work.

Just as the classical version of this equation encodes dilatation and special conformal invariance of the action, so~\eq{CERG} directly encodes the ERG equation~\eq{ERGE} and its special conformal partner~\eq{ERGE-SCT}. Solutions to~\eq{CERG} yield CFTs and generate the improvement term of the energy-momentum tensor upon substitution of~\eq{H} into either~\eq{Pol-condition-d>2} or~\eq{Pol-condition-d=2}.

\begin{acknowledgments}
	I thank Hugh Osborn for comments on the manuscript. Special thanks go to the anonymous referee  for reading the paper so carefully and for providing much constructive feedback.
\end{acknowledgments}

\appendix

\section{ERG Conventions}
\label{app:ERG}

Whereas, in this paper, we work with the full Wilsonian effective action, $\Stot$, more usually the ERG equation is phrased in terms terms of $\Sint$, defined via:
\be
	\Stot = \hf \dfield \cdot \ep^{-1} \cdot \dfield + \Sint
.
\ee
In terms of $\Sint$, the fixed-point ERG equation and its partner expressing special conformal invariance are:
\begin{subequations}
\begin{align}
	\biggl\{
		\dil{\delta} \dfield \cdot \fder{}{\dfield}
		+ \hf \fder{}{\dfield} \cdot G \cdot \fder{}{\dfield}
		-\frac{\eta}{2} \dfield \cdot \ep^{-1} \cdot \dfield
	\biggr\}
	e^{-\Sint}
&
	=
	0,
\label{eq:ERG-Int}
\\
	\biggl\{
		\sct{\delta}{\mu} \dfield \cdot \fder{}{\dfield}
		+ \hf \fder{}{\dfield} \cdot G_\mu \cdot \fder{}{\dfield}
		-\frac{\eta}{2} \dfield \cdot {\ep^{-1}}_\mu \cdot \dfield
		- \eta \partial_\alpha \dfield \cdot \cutoff^{-1} \cdot G \cdot 				\fder{}{\dfield}
	\biggr\}
	e^{-\Sint}
&
	=
	0
,
\label{eq:SCT-Int}
\end{align}
\end{subequations}
with, recalling~\eq{G_mu},
\be
	{\ep^{-1}}_\mu(x,y) = (x+y)_\mu \ep^{-1}(x,y)
.
\ee
We now show how to bring these equations into the form utilized in the rest of the paper. Starting with~\eq{ERG-Int}, observe that:
\begin{align}
	e^{-\hf \dfield \cdot \ep^{-1} \cdot \dfield}
	\biggl[
		\dil{\delta} \dfield \cdot \fder{}{\dfield}
	,
		e^{\hf \dfield \cdot \ep^{-1} \cdot \dfield}
	\biggr]
&
	=
	-
	\hf
	\dfield
	\cdot
		\bigl(
			\dil{\D-\delta} \ep^{-1} + \ep^{-1} \dilL{\D-\delta}
		\bigr)
	\cdot
	\dfield
\nonumber
\\
	&
	=
	\hf 
	\dfield 
	\cdot 
	\bigl(
		\eta \ep^{-1} - \ep^{-1} \cdot G \cdot \ep^{-1}
	\bigr)
	\cdot 
	\dfield	
\end{align}
and also:
\be
	e^{-\hf \dfield \cdot \ep^{-1} \cdot \dfield}
	\biggl[
		\hf \fder{}{\dfield} \cdot G \cdot \fder{}{\dfield}
	,
		e^{\hf \dfield \cdot \ep^{-1} \cdot \dfield}
	\biggr]
	=
	\dfield \cdot \ep^{-1} \cdot G \cdot \fder{}{\dfield}
	+
	\hf
	\dfield
	\cdot
		\ep^{-1} \cdot G \cdot \ep^{-1}
	\cdot
	\dfield
\ee
where, consistent with the rest of this paper, we have ignored a vacuum term on the \rhs.
It is thus apparent that~\eq{ERG-Int} transforms into~\eq{ERGE}.

To process~\eq{SCT-Int} we exploit the following result for some $U\bigl((x-y)^2\bigr)$ and some $V\bigl((x-y)^2\bigr)$. If we suppose that
\be
		\dil{\Delta} U + U \dilL{\Delta} = V
\ee
then, defining
\be
	V_\mu(x,y) = (x+y)_\mu V\bigl((x-y)^2\bigr)
,
\label{eq:V_mu}
\ee
 it follows that
\be
	\sct{\Delta}{\mu} U + U \sctL{\Delta}{\mu} = V_\mu.
\ee
Recalling~\eq{K(x,y)}, we see that
\be
	e^{-\hf \dfield \cdot \ep^{-1} \cdot \dfield}
	\biggl[
		\sct{\delta}{\mu} \dfield \cdot \fder{}{\dfield}
	,
		e^{\hf \dfield \cdot \ep^{-1} \cdot \dfield}
	\biggr]
	=
	\hf 
	\dfield 
	\cdot 
	\bigl(
		\eta \ep^{-1} - \ep^{-1} \cdot G \cdot \ep^{-1}
	\bigr)_\mu
	\cdot 
	\dfield
,
\ee
where the subsrcipt $\mu$ is to be interpreted as in~\eq{V_mu}.
The second result we require is
\be
	e^{-\hf \dfield \cdot \ep^{-1} \cdot \dfield}
	\biggl[
		\hf \fder{}{\dfield} \cdot G_\mu \cdot \fder{}{\dfield},
	,
		e^{\hf \dfield \cdot \ep^{-1} \cdot \dfield}
	\biggr]
	=
	\dfield \cdot \ep^{-1} \cdot G_\mu \cdot \fder{}{\dfield}
	+
	\hf
	\dfield
	\cdot
		\ep^{-1} \cdot G_\mu \cdot \ep^{-1}
	\cdot
	\dfield
.
\ee
Next, we combine various terms:
\be
	\dfield
	\cdot
	\bigl(
		\ep^{-1} \cdot G_\mu \cdot \ep^{-1} - \bigl( \ep^{-1} \cdot G \cdot \ep^{-1} \bigr)_\mu
	\bigr)
	\cdot
	\dfield
	=
	\dfield
	\cdot
	\bigl(
		\bigl[\ep^{-1}, X_\mu \bigr] \cdot G \cdot \ep^{-1}
		- \ep^{-1} \cdot G \cdot \bigl[\ep^{-1}, X_\mu \bigr]
	\bigr)
	\cdot
	\dfield
	= 0
,
\label{eq:odd}
\ee
where $\bigl[\ep^{-1}, X_\mu \bigr](x,y) = (x-y)_\mu \ep^{-1} \bigl((x-y)^2\bigr) = \partial_\mu F\bigl((x-y)^2\bigr)$, for some $F$. The presence
of $\partial_\mu$ means that each of the two terms in the middle step of~\eq{odd} are separately odd, and so the integrals vanish. Putting everything together reproduces~\eq{ERGE-SCT}.

\section{Standard form of Key Results}
\label{app:uncondensed}

Recall that it was convenient to split the energy-momentum tensor into classical and quantum pieces, as in~\eq{split}.
Each of these was then separately decomposed along the lines of~\eqs{Tbar}{TbarSym}. For the quantum piece, the latter had explicit contributions given by~\eqs{QbarSym}{F_lab}, which we now unpack into standard notation, producing
\begin{multline}
	\symmetric{\conserved{\QEMT}}_{\alpha \beta}(x) e^{-\Stot}
	=
	\Int{y}\Int{z}
	\biggl\{
		\hf \delta_{\alpha \beta} \fder{}{\dfield(y)}	
		\cutoff(y,x)\R(x,z) \fder{}{\dfield(z)}
\\
		+
		\Int{u}
		\biggl(
			2\partial_\alpha \partial_\beta \dfield(u)
			\cutoff^{-1}(u,y)
			\core{\cutoff}(y,z;x)
			\fder{}{\dfield(z)}
\\
			-
			\fder{}{\dfield(u)} 
			\cutoff(u,y)
			\core{\R}(y,z;x)
			\partial_\alpha \partial_\beta \fder{}{\dfield(z)}
			+
			\partial_\alpha \partial_\beta \fder{}{\dfield(y)}
			\core{\cutoff}(y,z;x)
			R(z,u)\fder{}{\dfield(u)}
		\biggr)
	\biggr\}
	e^{-\Stot}
\end{multline}
and
\begin{multline}
	F_{\lambda \alpha \beta}(x) e^{-\Stot}
	= 
	-\delta_{\lambda \alpha}
	\Int{y}\Int{z}\Int{u}
	\biggl(
		 \partial_\beta \dfield(u)
		 \cutoff^{-1}(u,y)
		 \core{K}(y,z;x)
		 \fder{}{\dfield(z)}
\\
		- 
		\hf \fder{}{\dfield(u)} 
		\cutoff(u,y)
		\core{\R}(y,z;x)
		\partial_\beta \fder{}{\dfield(z)}
		+ 
		\hf \partial_\beta \fder{}{\dfield(y)}
		\core{\cutoff}(y,z;x)
		\R(z,u) \fder{}{\dfield(u)}
	\biggr)
	e^{-\Stot} 
.
\end{multline}
These two terms are sufficient to reconstruct the quantum part of the energy-momentum tensor, $\QEMT_{\alpha\beta}$, up to the additional pieces required for tracelessness. In turn, the latter are determined via~\eqs{Pol-condition-d>2}{Pol-condition-d=2}, given some $H_{\tau\lambda}$ (the ERG representation of $\arbitrary{H}_{\tau\lambda}$). $H_{\tau\lambda}$ was split into classical and quantum pieces, according to~\eq{H=h+I}. The quantum piece, $I_{\alpha \beta}$, may be determined via two steps. First, the trace $\QEMT_{\alpha\alpha}$ was constructed using the recipe~\eq{traceOfEMT} to yield~\eq{Q_aa}, reproduced here in standard notation:
\begin{multline}
	\QEMT_{\alpha\alpha}(x) 
	= \partial_\lambda \partial_\tau I_{\tau \lambda}(x)
	+
	e^{\Stot}
	\Int{y}\Int{z}
	\biggl\{
		\frac{\D}{2}
		\fder{}{\dfield(y)} 
		\cutoff(y,x) \times \R(x,z)
		\fder{}{\dfield(z)}
\\
		\Int{u}
		\biggl(
			-\partial_\lambda \dfield(u) 
			\cutoff^{-1}(u,y)
			\bigl[
				\core{K}_\lambda
				+(1-\D)\partial_\lambda\core{K}
			\bigr](y,z;x)
			\fder{}{\dfield(z)}
\\
			-\hf \fder{}{\dfield(u)}
			\cutoff(u,y)
			\bigl[
				\core{R}_\lambda
				+(\D-1) \partial_\lambda \core{R}
			\bigr](y,z;x)
			\partial_\lambda \fder{}{\dfield(z)}
\\
			-\hf 
			\partial_\lambda \fder{}{\dfield(y)}
			\bigl[
				\core{\cutoff}_\lambda
				+(1-\D) \partial_\lambda \core{\cutoff}
			\bigr](y,z;x)
			\R(z,u)
			\fder{}{\dfield(u)}
	\biggr\}
	e^{-\Stot}
.
\end{multline}
Secondly, this may be compared with the trace of the energy-momentum tensor as given by~\eq{T_aa-ERG} from which the quantum piece can be extracted by using~\eq{extract}:
\begin{multline}
	\QEMT_{\alpha\alpha}(x)
	=
	\delta e^{\Stot}
	\Int{y}\Int{z}
	\biggl\{
		\fder{}{\dfield(y)} \cutoff(y,x) \R(x,z) \fder{}{\dfield(z)}
\\
		-\pder{}{x_\lambda}
		\Int{u}
		\dfield(u) \cutoff^{-1}(u,y)
		\core{\cutoff}_\lambda(y,z;x)
		\fder{}{\dfield(z)}		
	\biggr\}
	e^{-\Stot}
.
\end{multline}
Thus, given a Wilsonian effective action, the energy-momentum tensor can, in principle, be constructed. However, by manipulating the consistency equation for the trace, a new equation, \eq{CERG},  was derived. Without exploiting condensed notation for the integrals, but utilizing~\eq{condensed_partials}, this equation may be written as
\begin{multline}
	-\partial_\lambda \partial_\tau \remainder{H}_{\tau\lambda}(x)
	 = 
	\biggl\{
		\D\Ltot(x)
		-
		\sum_{i=1}^\infty
		\bigl[ \partialprod{\sigma}{1}{i}, x\cdot\partial \bigr] \dfield(x)
		\pder{\Ltot(x)}{(\partialprod{\sigma}{1}{i}\dfield(x))}
		+
		\delta \dfield(x) \fder{}{\dfield(x)}
\\
		+ 
		\hf
		\Int{y} \Int{z}
		\biggl(
		\dfield(y)
		\biggl(
			\ep^{-1}(y,z) G(z,x) \fder{}{\dfield(x)}
			+ \ep^{-1}(y,x) G(x,z) \fder{}{\dfield(z)}
		\biggr)
	\\
		+ 
		\hf
		\Int{y} 
		\fder{}{\dfield(y)} G(y,x)\fder{}{\dfield(x)}
\\
	+\pder{}{x_\lambda}
	\biggl[
		\hf
		\bigl(
			\delta_{\omega \lambda} \delta_{\rho\sigma}
			- 2\delta_{\omega \rho} \delta_{\sigma\lambda}
		\bigr)
		\sum_{i=2}^\infty
		\bigl[
			\bigl[
				\partialprod{\sigma}{1}{i}, x_\sigma
			\bigr],
			x_\rho \partial_\omega
		\bigr]
		\dfield(x)
		\pder{\Ltot(x)}{(\partialprod{\sigma}{1}{i} \dfield(x))}
\\
		+
		\frac{\eta}{4}
		\Int{y}\Int{z}
		\partial_\lambda \dfield(y)
		\biggl(
			\cutoff^{-1}(y,z) G(z,x) \fder{}{\dfield(x)}
			+
			\cutoff^{-1}(y,x) G(x,z) \fder{}{\dfield(z)}
		\biggr)
	\biggr]
	\biggr\}
	e^{-\Stot}
.
\end{multline}

\end{document}